%% file: main.tex
\PassOptionsToPackage{switch,pagewise,columnwise,mathlines}{lineno}
\pdfoutput=1
\documentclass[fleqn,usenatbib]{mnras}

\usepackage{newtxtext,newtxmath} 

\usepackage[T1]{fontenc}
\usepackage{ae,aecompl}

\usepackage{etoolbox}
\usepackage[table,xcdraw]{xcolor}
\selectcolormodel{cmyk}
\usepackage{graphicx}
\DeclareGraphicsExtensions{.pdf,.jpg,.png,.eps}
\graphicspath{{figs/}}
\usepackage{amsmath}  
\usepackage{mathtools}  
\usepackage{amsfonts}
\usepackage{booktabs}  
\usepackage{multirow}
\usepackage{caption}
\usepackage{subcaption}
\usepackage{etoolbox}
\usepackage{siunitx}
\usepackage{mwe,tikz}
\usepackage[percent]{overpic}
\usepackage{comment}
\usepackage{xurl}

\usepackage[finalizecache=false,frozencache=true,cachedir=minted-cache]{minted}

\newcommand{\nbpix}{N}
\newcommand{\nfourier}{K}
\newcommand{\nband}{L}

\newcommand{\nfacet}{Q}

\newcommand{\nblock}{B}

\newcommand{\ndata}{M}
\newcommand{\ncube}{C}

\newcommand{\bs}[1]{\boldsymbol{#1}}

\newcommand{\minimize}[2]{\ensuremath{\underset{\substack{{#1}}}{\mathrm{minimize}}\;\;#2 }}

\DeclarePairedDelimiter{\norm}{\lVert}{\rVert}

\DeclareMathOperator{\runp}{\texttt{run}_\textsubscript{\texttt{pips}}}
\DeclareMathOperator{\run}{\texttt{run}}
\DeclareMathOperator{\cpup}{\texttt{cpu}_\textsubscript{\texttt{pips}}}
\DeclareMathOperator{\cpu}{\texttt{cpu}}

\definecolor{darkgreen}{cmyk}{0.8,0,0.8,0.45}
\definecolor{lightgreen}{cmyk}{0.8,0,0.8,0.25}
\usepackage[ruled,vlined,linesnumbered]{algorithm2e}

\SetCommentSty{mycommfont}
\SetKwInput{Parameter}{Parameters} 
\let\oldnl\nl
\newcommand{\nonl}{\renewcommand{\nl}{\let\nl\oldnl}} 

\usepackage{tikz}
\usetikzlibrary{shapes,arrows,positioning,spy,quotes,arrows.meta}

\usepackage[switch,pagewise,columnwise,mathlines]{lineno} 
\title[Faceted HyperSARA]{Parallel faceted imaging in radio interferometry via proximal splitting (Faceted HyperSARA): II. Code and real data proof of concept}

\author[Thouvenin et al.]{Pierre-Antoine Thouvenin,$^{1,2}$
Arwa Dabbech,$^{1}$
Ming Jiang,$^{3,4}$
Abdullah Abdulaziz,$^{1}$
Jean-Philippe Thiran,$^{3}$
\newauthor Adrian Jackson,$^{5}$ and Yves Wiaux$^{1}$\thanks{E-mail: y.wiaux@hw.ac.uk}
\\
$^{1}$Institute of Sensors, Signals and Systems, Heriot-Watt University, Edinburgh EH14 4AS, United Kingdom\\
$^{2}$Universit{\'e} de Lille, CNRS, Centrale Lille, UMR 9189 CRIStAL, F-59000 Lille, France\\
$^{3}$Signal Processing Laboratory 5 (LTS5), \'Ecole Polytechnique F\'ed\'erale de Lausanne, CH-1015, Lausanne, Switzerland\\
$^{4}$National Laboratory of Radar Signal Processing, Xidian University, Xi'an 710071, China\\
$^{5}$EPCC, University of Edinburgh, Edinburgh EH8 9BT, United Kingdom
}

\date{Accepted XXX. Received YYY; in original form ZZZ}

\pubyear{2022}

\begin{document}
\label{firstpage}
\pagerange{\pageref{firstpage}--\pageref{lastpage}}
\maketitle

\begin{abstract}
In a companion paper, a faceted wideband imaging technique for radio interferometry, dubbed Faceted HyperSARA, has been introduced and validated on synthetic data. Building on the recent HyperSARA approach, Faceted HyperSARA leverages the splitting functionality inherent to the underlying primal-dual forward-backward algorithm to decompose the image reconstruction over multiple spatio-spectral facets. The approach allows complex regularization to be injected into the imaging process while providing additional parallelization flexibility compared to HyperSARA. The present paper introduces new algorithm functionalities to address real datasets, implemented as part of a fully fledged MATLAB imaging library made available on \href{https://basp-group.github.io/Puri-Psi/}{\textsc{Github}}. A large scale proof-of-concept is proposed to validate Faceted HyperSARA in a new data and parameter scale regime, compared to the state-of-the-art. The reconstruction of a 15~GB wideband image of Cyg~A from 7.4~GB of VLA data is considered, utilizing 1440 CPU cores on a HPC system for about 9 hours. The conducted experiments illustrate the reconstruction performance of the proposed approach on real data, exploiting new functionalities to leverage known direction-dependent effects (DDEs), for an accurate model of the measurement operator, and an effective noise level accounting for imperfect calibration. They also demonstrate that, when combined with a further dimensionality reduction functionality, Faceted HyperSARA enables the recovery of a 3.6~GB image of Cyg~A from the same data using only 91 CPU cores for 39 hours. In this setting, the proposed approach is shown to provide a superior reconstruction quality compared to the state-of-the-art wideband CLEAN-based algorithm of the WSClean software. 
\end{abstract}

\begin{keywords}
techniques: image processing, techniques: interferometric. 
\end{keywords}


\section{Introduction}\label{sec:intro}

The Faceted HyperSARA approach~\citep{Thouvenin2021} is a novel wideband imaging technique for radio interferometry, enabling a fully distributed reconstruction over the wideband data and image via data blocking and spatial and spectral faceting. The approach thus paves the way towards scalability to the large data and image sizes expected from modern radio telescopes. This paper presents a new MATLAB library implementing the Faceted HyperSARA approach for real radio interferometric (RI) data with a focus on (i) its structure for an enhanced user experience and (ii) the added functionalities for a fast and accurate RI measurement operator in the presence of unknown noise levels due to calibration errors. More specifically, models of the RI measurement operator can encapsulate the $w$-term, a direction-dependent effect resulting from the non-coplanarity of the radio array, via $w$-projection \citep{Cornwell2008,wolz2013,Dabbech2017}. Available direction-dependent effect solutions of other origins (\emph{e.g.}, atmospheric or instrumental) obtained from a calibration pre-processing step \citep{Repetti2017,Dabbech2021}, can also be integrated into the measurement operator. Furthermore, unknown effective noise levels induced by the imperfect calibration are estimated using an adaptive strategy inspired from \citet{Dabbech2018}. Finally, large data volumes can be accommodated via a visibility gridding-based data dimensionality reduction scheme \citep{Kartik2017}.

Two proof-of-concept experiments are proposed to assess the algorithm's performance on real wideband RI data acquired with the Very Large Array (VLA).
The first experiment is conducted to evaluate the reconstruction performance of Faceted HyperSARA in terms of imaging quality through the reconstruction of a 15~GB image cube of Cyg~A from 7.4~GB visibilities across 480 channels, spanning the frequency range 3.979--8.019~GHz . Utilizing the same data, a second experiment is performed with Faceted HyperSARA, combined with a joint image and a visibility gridding-based data dimensionality reduction technique to accommodate larger data volumes. More precisely, given the moderate spectral variation of Cyg~A in the frequency range of interest, we target a 16-fold reduction of the spectral resolution. The visibility gridding offers a 7-fold reduction in the data volume. 

The article is organized as follows. In Section~\ref{sec:background}, the faceted prior model and objective function underlying the Faceted HyperSARA approach are summarized. Newly added functionalities for a more robust and scalable measurement operator are detailed in Section~\ref{sec:newfeatures}. Section~\ref{sec:matlab} introduces a fully fledged MATLAB library, addressing the RI imaging problem with algorithms from the SARA family. These are the monochromatic SARA approach~\citep{Carrillo2012,Onose2017} and the wideband approaches HyperSARA~\citep{Abdulaziz2016,Abdulaziz2019} and Faceted HyperSARA. To study the performance of the latter approach, a first experiment is conducted utilizing VLA data in Section~\ref{sec:exp_real}, assessing its imaging quality and computing time in comparison with the SARA approach \citep{Onose2017}. Section~\ref{sec:exp_DR} illustrates the potential of combining Faceted HyperSARA and a dimensionality reduction technique to handle larger data and image sizes, utilizing the same data as in Section~\ref{sec:exp_real}. The approach is assessed in comparison with the state-of-the-art
wideband CLEAN-based method of the WSClean software~\citep{Offringa2017}. Finally, conclusions and perspectives are reported in Section~\ref{sec:conclusion}.

\section{Faceted HyperSARA: a summary} \label{sec:background}

The problem and notations used to formulate the RI imaging problem are recalled in this section (notation summary provided in Table~\ref{tab:notations}). A short summary of the Faceted HyperSARA approach, introduced in the companion paper~\citep[Sections 2 and 3]{Thouvenin2021}, is also provided. 

\input{tab_notations}

\subsection{Problem formulation} \label{ssec:background:imaging}

Wideband RI imaging refers to the reconstruction process of unknown 
images of the sky over $\nband$ frequency channels from noisy data -- referred to as \emph{visibilities} -- acquired by an array of antennas. Assuming the image to be reconstructed covers a small field of view on the celestial sphere with perfect antenna calibration, each visibility point, acquired by a pair of antennas at a given frequency, corresponds to a spatial Fourier component of the sky surface brightness. The associated spatial 
frequency (in units of the observation's wavelength) corresponds to the projection of the corresponding baseline onto the plane perpendicular to the line of sight~\citep{Thompson2007}. These
provide an incomplete coverage of the 2D spatial Fourier plane (also called $uv$-coverage) of the image. The measurement process can be modeled for each frequency channel $l \in \{1, \dotsc, \nband \}$ as~\citep{Abdulaziz2016,Abdulaziz2019}
\begin{linenomath}
\begin{equation} \label{eq:model}
 \mathbf{y}_l = \bs{\Phi}_l \overline{\mathbf{x}}_l + \mathbf{n}_l, \: \text{with } \bs{\Phi}_l = \bs{\Theta}_l \mathbf{G}_l \mathbf{FZ},
\end{equation}
\end{linenomath}
where $\mathbf{y}_l \in \mathbb{C}^{\ndata_l}$ is the vector of the whitened visibilities (a process known as natural weighting in RI imaging) acquired at the frequency channel indexed by $l~\in~\{ 1, \dotsc, \nband \}$, $\overline{\mathbf{x}}_l \in \mathbb{R}_+^\nbpix$ denotes the underlying unknown image, 
and $\mathbf{n}_l \in \mathbb{C}^{\ndata_l}$ represents the measurement noise modeled as a realization of a standard complex independent and identically distributed random variable. The measurement operator $\bs{\Phi}_l$ represents a non-uniform fast Fourier transform (NUFFT)~\citep{Fessler2003}, composed of a zero-padding and scaling operator $\mathbf{Z} \in \mathbb{R}^{\nfourier \times \nbpix}$ (with $\nfourier$ the number of points after zero-padding), the 2D Discrete Fourier Transform represented by the matrix $\mathbf{F} \in \mathbb{C}^{\nfourier \times \nfourier}$, and an interpolation matrix $\mathbf{G}_l \in \mathbb{C}^{\ndata_l \times \nfourier}$ whose rows contain a compact-support interpolation kernel centered at the corresponding $uv$-point. Note that available models of direction-dependent effects can be incorporated in the matrix $\mathbf{G}$ as convolution with their compact-support Fourier representation \citep{Repetti2017, Dabbech2021}. The diagonal noise whitening matrix $\bs{\Theta}_l$ contains the inverse of the noise standard deviation associated with the raw measurements.

To efficiently address the large number of visibilities in the reconstruction algorithm, the visibility vectors $\mathbf{y}_l$ can be further decomposed into $\nblock$ blocks, $\mathbf{y}_l = (\mathbf{y}_{l,b})_{1 \leq b \leq \nblock}$~\citep{Onose2016, Abdulaziz2019}, where $\mathbf{y}_{l,b} \in \mathbb{C}^{\ndata_{l,b}}$ is composed of $\ndata_{l,b}$ visibilities. The corresponding decomposition of the measurement operator is hereafter denoted by $\bs{\Phi}_{l} = (\bs{\Phi}_{l,b})_{1 \leq b \leq \nblock}$.

\subsection{Faceted HyperSARA} \label{ssec:background:fhs}
 
\subsubsection{Optimization problem} \label{sssec:background:fhs:problem}

From the perspective of optimization theory, reconstructing the original wideband image $\overline{\mathbf{X}} = (\overline{\mathbf{x}}_l)_{1 \leq l \leq \nband}$ from incomplete noisy Fourier measurements can be formulated as the resolution of the following constrained problem
\begin{linenomath}
\begin{equation} \label{eq:problem}
	\minimize{\mathbf{X} = (\mathbf{x}_l)_{1 \leq l \leq \nband} \in \mathbb{R}_+^{\nbpix \times \nband}} \sum_{l =1}^\nband \sum_{b =1}^\nblock \iota_{\mathcal{B}(\mathbf{y}_{l,b}, \varepsilon_{l,b})} \bigl( \bs{\Phi}_{l,b}\mathbf{x}_{l} \bigr) + r(\mathbf{X}),
\end{equation}
\end{linenomath}
where the indices $(l,b) \in \{ 1, \dotsc , \nband \} \times \{ 1, \dotsc , \nblock \}$ identify the data block $b$ of channel $l$, and the function $\iota_{\mathcal{B}(\mathbf{y}_{l,b},\varepsilon_{l,b})}$ is the indicator function of the set $\mathcal{B}(\mathbf{y}_{l,b},\varepsilon_{l,b}) = \bigl \{ \mathbf{z} \in \mathbb{C}^{\ndata_{l,b}} \mid \norm{\mathbf{z} - \mathbf{y}_{l,b}}_2 \leq \varepsilon_{l,b} \bigr\}$, the $\ell_2$-ball centred in $\mathbf{y}_{l,b}$ of radius $\varepsilon_{l,b} > 0$\footnote{Let $\mathcal{C}$ be a non-empty, closed, convex subset of $\mathbb{C}^\nbpix$, then $\iota_{\mathcal{C}}$ denotes the indicator function of $\mathcal{C}$, defined by $\iota_{\mathcal{C}} (\mathbf{z}) = +\infty$ if $\mathbf{z} \in \mathcal{C}$ and 0 otherwise.}. Assuming perfect calibration, the parameter $\varepsilon_{l,b}$ is such that $\varepsilon_{l,b}^2 = \varepsilon_{l}^2 \ndata_{l,b} / \ndata_l$, with $\varepsilon_{l}^2 = \ndata_{l} + 2 \sqrt{\ndata_{l}}$, reflecting the statistics of the noise in the corresponding data block, assuming a Chi-squared distribution~\citep[see][for further details]{Carrillo2012,Onose2016}. The function $r$ denotes a regularization term. Finally, a non-negativity constraint is imposed on the wideband image, as it corresponds to an intensity image.

In the minimized objective function, the indicator functions $\iota_{\mathcal{B}(\mathbf{y}_{l,b}, \varepsilon_{l,b})}$ act as \emph{data-fidelity} terms, ensuring the consistency of the modeled visibilities with the measurements~\citep{Carrillo2012}, whereas the function $r$ encodes a prior model for the unknown image cube to address the ill-posedness of the inverse problem described in~\eqref{eq:model}. Note that the prior function involves regularization parameters, which are set from the noise level approximated through the heuristics proposed in \citet{Thouvenin2021}. Both the HyperSARA and Faceted HyperSARA approaches rely on a low-rankness and average joint-sparsity prior (see \citep[Section 2]{Thouvenin2021}), whereas the SARA approach relies on a sparsity prior. The Faceted HyperSARA approach is summarized in the next paragraph.

\subsubsection{Faceted HyperSARA approach} \label{sssec:background:fhs:faceting}

Building on the original HyperSARA approach, Faceted HyperSARA relies on the two distinct levels of faceting described below.

\paragraph*{Spectral faceting.} The wideband image cube can be reconstructed as a collection of $C$ image sub-cubes, each recovered from a subset of the frequency channels and leveraging an independent model prior. In \citep{Thouvenin2021}, channel-interleaved spectral sub-cubes have been considered to ensure that each imaging problem can exploit the spectral information over the extent of the observed frequency band. In this setting, a collection of $\ncube$ problems are solved, where the data-fidelity terms are frequency channel-based and the regularization terms acts on the image sub-cubes separately. Solving the overall objective function involved in~\eqref{eq:problem} reduces to solving the sum of independent objective functions associated with the $C$ sub-cubes as follows
\begin{linenomath}
\begin{multline} \label{eq:problem-split-subcube}
	\minimize{\mathbf{X}_c \in \mathbb{R}_+^{\nbpix \times \nband_c}} 
	\sum_{l=1}^{\nband_c} \sum_{b=1}^\nblock \iota_{\mathcal{B}(\mathbf{y}_{c,l,b}, \varepsilon_{c,l,b})} \bigl( \bs{\Phi}_{c,l,b}\mathbf{x}_{c,l} \bigr)
	+ r_c (\mathbf{X}_c),
\end{multline}
\end{linenomath}
where $c \in \{1, \dotsc, \ncube\}$ and $\nband_c$ denotes the number of channels of $c$-th sub-cube. For each channel $l \in \{1, \ldots, \nband_c\}$ and data-block $b$, $\mathbf{y}_{c,l,b} \in \mathbb{C}^{\ndata_{c,l,b}}$ denotes the vector of $\ndata_{c,l,b}$ visibilities, $\bs{\Phi}_{c,l,b}$ the associated measurement operator, and $\varepsilon_{c,l,b}$ the $\ell_2$ ball radius.
Finally, the matrix $\mathbf{X}_c = (\mathbf{x}_{c,l})_{1 \le l \le \nband_c} \in \mathbb{R}^{\nbpix \times \nband_c}$ is the $c$-th sub-cube of the full image cube $\mathbf{X} = (\mathbf{X}_c)_{1 \leq c \leq \ncube}$, with $\mathbf{x}_{c,l} \in \mathbb{R}^\nbpix$ the $l$-th image of the sub-cube $\mathbf{X}_c$. The regularization term $r_c \colon \mathbb{R}^{\nbpix \times \nband_c} \to ]-\infty, +\infty]$ acts on the $c$-th image sub-cube. In addition to non-negativity, it enforces low-rankness and average joint-sparsity of the image sub-cube, encapsulated in non-convex log-sum functions \citep[see ][Section 2.3, for full details]{Thouvenin2021}.

\paragraph*{Spatial faceting.} The regularization term $r_c$ from~\eqref{eq:problem-split-subcube} can be further decomposed into a sum of $\nfacet$ spatial facet-specific terms, \emph{i.e.}, exclusively affecting contiguous pixels. On the one hand, the wavelet dictionaries involved in the average joint-sparsity prior from HyperSARA can be exactly decomposed over overlapping facets~\citep[Chapter 4]{Prusa2012}. The associated regularization term can thus be reformulated as a sum of several per-facet terms. On the other hand, a faceted low-rankness prior is considered to provide a more tractable alternative to the global low-rankness prior from HyperSARA involving the singular value decomposition (SVD) of the full image cube.

On a final note, the HyperSARA approach can be seen as a particular case of the Faceted HyperSARA approach with $(\ncube, \nfacet) = (1, 1)$.

\subsubsection{Algorithm} \label{sssec:background:fhs:algorithm}

To efficiently solve~\eqref{eq:problem-split-subcube} and address its non-convex regularization terms, a reweighting iterative scheme~\citep{Candes2009} relying on the primal-dual forward-backward (PDFB) algorithmic structure~\citep{Condat2013, Vu2013, Pesquet2014} is considered \citep[see][Section 3]{Thouvenin2021}. More precisely, the \emph{outer} reweighting algorithm allows the resolution of the non-convex problem via the resolution of a sequence of convex sub-problems
\citep[see][Section 3.1, Algorithm 1]{Thouvenin2021}. These sub-problems are solved efficiently using PDFB as \emph{inner} algorithmic structure, where full parallelization of all block-specific data-fidelity terms and facet-specific regularization terms is achieved~\citep[see][Section 3.2, Algorithm 2]{Thouvenin2021}.

Compared to HyperSARA, Faceted HyperSARA thus offers more parallelization flexibility in that:
\begin{enumerate}
    \item spectral faceting can be used to independently reconstruct the wideband image in parallel over non-overlapping groups of channels, which lead to $\ncube$ independent optimization problems of the form~\eqref{eq:problem-split-subcube}, formulated for a single wideband sub-cube $\mathbf{X}_c$. This feature can be used to exploit more computational resources for the outer reweighting algorithm;
    \item spatial faceting allows a larger number of workers to be involved in the inner PDFB algorithm underlying the approach \emph{via} splitting. Furthermore, per-facet regularization terms act on a smaller part of the wideband image, and can be assigned to workers on a per-facet basis. Further details about the communications between the workers involved in the PDFB algorithm are provided in~\citet[Section 3.3]{Thouvenin2021}.
\end{enumerate}

\section{Towards an accurate and scalable measurement operator} \label{sec:newfeatures} 

In this section, we describe new functionalities compared to~\citep{Thouvenin2021}. These are aimed at (i) enhancing the accuracy of the RI measurement operator by incorporating available models of the DDEs and correcting for the $w$-term in the context of wide-field imaging, (ii) addressing the limitations of the imperfect calibration and its impact on the noise level estimate and (iii) improving the scalability potential of the measurement operator via data dimensionality reduction. These features are crucial to achieve high precision imaging from large data volumes.

\subsection{Accounting for estimates of the direction-dependent effects} \label{ssec:newfeatures:dde}

Based on a compact-support model in the spatial Fourier domain, DDE estimates can be accounted for in the measurement operator when building the de-gridding matrix $\mathbf{G}$~\citep{Repetti2017}. 
More precisely, each row of $\mathbf{G}$, associated with a data point, results from a linear convolution between the compact Fourier representation of the DDEs of the associated antenna pair and a non-uniform Fourier interpolation kernel~\citep{Dabbech2021}. The particular case of the direction independent effects (DIE), where the antenna gains are modeled via complex values, is addressed in the imaging by applying the available DIE solutions directly to the data.

\subsection{Correction of the $w$-term via $w$-projection} \label{ssec:newfeatures:w}

The non-coplanarity of the radio array results in variations in the projections of the baselines onto the direction of observation, defining their $w$-coordinates. The so called $w$-term -- a phase modulation in the image domain -- becomes non-negligible, especially for wide-field imaging \citep{Cornwell2004}. To take this geometric DDE into account in the measurement operator, we consider the $w$-projection approach \citep{Cornwell2008}. Each visibility is modeled from the Fourier transform of the radio map using a linear convolution with the spatial Fourier transform of the $w$-term.
Building on \citet{wolz2013,Dabbech2017}, we consider compact $w$-terms with adaptive resolution based on their estimated spatial bandwidth. The resulting $w$-kernels are further sparsified via adaptive sparsification procedures. The kernels are incorporated into the de-gridding matrix $\mathbf{G}$ via a linear convolution with the non-uniform Fourier interpolation kernel, resulting in accurate and fast measurement operators.

\subsection{Effective noise level estimation}
\label{ssec:newfeatures:noise}

Given the implication of the noise level in the definition of the minimization problem through (i) the $\ell_2$ bounds $(\varepsilon_{l,b})_{(1 \leq l \leq \nband, 1 \leq b \leq \nblock)}$ involved in the data-fidelity constraints in~\eqref{eq:problem}, and (ii) the regularization parameters resulting from the heuristics proposed in \citet{Thouvenin2021}, it is of paramount importance to provide reliable noise estimates to the intensity imaging algorithm, to ensure its effectiveness in terms of reconstruction quality.
In the case of perfect calibration, the standard deviation of the naturally-weighted data is assumed to be equal to one. Yet, in practice, calibration errors tend to compromise the accuracy the measurement operator, resulting into an effective noise level higher than the theoretical instrumental noise, especially in the context of the highly sensitive modern telescopes. In this case, estimates of the effective noise can be obtained by (i) relying on model visibilities obtained from a calibration pre-processing step, where the noise level per each channel is derived as the standard deviation of the associated residual visibilities or (ii) activating a new functionality for adaptive estimation of the effective noise level within a few PDFB iterations inspired by \citet{Dabbech2018,Abdulaziz2019}. Starting from inactive image priors, the strategy consists in progressively updating the noise level estimates and associated parameters (i.e. $\ell_2$ bounds and regularization parameters), throughout the iterative process. Once the $\ell_2$ bounds are satisfied, the noise-related parameters are fixed, thus defining the final problem to be solved, and the intensity imaging algorithm is run to convergence.

\subsection{Data dimensionality reduction}
\label{ssec:newfeatures:gridding}

To handle large data volumes, the RI data and, consequently, the associated measurement operator, 
can be embedded into a lower-dimensional sub-space, especially when the number of measurements in each channel exceeds the spatial dimension of the target model cube. \citet{Kartik2017} proposed various data dimensionality reduction schemes. In this paper, we consider a visibility gridding-based scheme due to its simplicity. An overview of the technique is given in the following lines in the context of Faceted HyperSARA. 

Consider $C$ independent channel-interleaved sub-problems, and the linear embedding operators $\bs{\Gamma}_{c,l,b}\in\mathbb{C}^{{\ndata}^\prime_{c,l,b}\times {\ndata}_{c,l,b}}$. The forward model~\eqref{eq:model} after data gridding reads
\begin{linenomath}
    \begin{equation}\label{newIP} 
        \mathbf{y}^\prime_{c,l,b} = \bs{\Phi}^\prime_{c,l,b} \mathbf{x}_l + \mathbf{n}^\prime_{c,l,b}, 
    \end{equation}
\end{linenomath}
where $\bs{\Phi}^\prime_{c,l,b} = \bs{\Gamma}_{c,l,b}\bs{\Phi}_{c,l,b}$, and $\mathbf{y}^\prime_{c,l,b}=\bs{\Gamma}_{c,l,b}\mathbf{y}_{c,l,b}\in\mathbb{C}^{\ndata^\prime_{c,l,b}}$ and $\mathbf{n}^\prime_{c,l,b}=\bs{\Gamma}_{c,l,b}\mathbf{n}_{c,l,b}\in\mathbb{C}^{\ndata^\prime_{c,l,b}}$ are the gridded data and noise vectors, respectively. In this formulation, the resulting measurement operator $\bs{\Phi}^\prime_{c,l,b}\in \mathbb{C}^{{\ndata}^\prime_{c,l,b}\times\nbpix}$ becomes blind to the original data dimension ${\ndata}_{c,l,b}$, and the reduced data dimension ${\ndata}^\prime_{c,l,b}$ can be significantly smaller than both the original data and image dimensions (${\ndata}_{c,l,b}, \nbpix$). The gridding operator reads $\mathbf{\Gamma}_{c,l,b} = \bs{\Lambda}_{l,b,c}\mathbf{G}_{c,l,b}^\dagger \bs{\Theta}^\dagger_{c,l,b}$,
with the diagonal matrix $\mathbf{\Lambda}_{c,l,b}\in\mathbb{R}^{{\ndata}^\prime_{c,l,b}\times {\ndata}^\prime_{c,l,b}}$ acting as a noise-whitening operator. 
The resulting measurement operator $\bs{\Phi}_{c,l,b}^\prime$ explicitly reads
\begin{linenomath}
    \begin{equation}\label{eq:red_phi_}
        \bs{\Phi}_{c,l,b}^\prime = \mathbf{\Lambda}_{c,l,b}{\mathbf{H}}_{c,l,b}\mathbf{FZ},
    \end{equation}
\end{linenomath}
that is encapsulating the sparse holographic matrix given by $\mathbf{H}_{c,l,b} = \mathbf{G}_{c,l,b}^\dagger \bs{\Theta}^\dagger_{c,l,b} \bs{\Theta}_{c,l,b} \mathbf{G}_{c,l,b} \in \mathbb{C}^{{\ndata}^\prime_{c,l,b} \times {\ndata}^\prime_{c,l,b}}$. The diagonal matrix $\bs{\Lambda}_{c,l,b}$ acts effectively on the nonzero rows of the matrix $\mathbf{H}_{c,l,b}$, with its entries corresponding to the inverse of the square root of the nonzero diagonal elements of the matrix $\mathbf{H}_{c,l,b}$ and zero otherwise. 
On the one hand, this form of weighting ensures that the embedded noise is approximately an independent and identically distributed Gaussian noise. On the other hand, it operates as a uniform weighting, as the diagonal elements of the holographic matrix encapsulate the density of the sampling of the gridded visibilities~\citep{Kartik2017}.

\section{MATLAB library} \label{sec:matlab}

In this section, we describe our proposed MATLAB library, that is focused on RI wideband intensity imaging. The library offers a collection of utility functions and scripts from data extraction from an RI measurement set (MS Table) to the reconstruction of a wideband intensity image over the field of view and frequency range of interest.

\noindent
The library is composed of three modules associated with (i) the formation of the measurement operator, including the new functionalities presented in Section~\ref{sec:newfeatures}, (ii) a faceted implementation of the sparsity dictionary involved in the average sparsity prior of the SARA family, and (iii) the imaging pipeline which enables interfacing with an RI measurement set and wideband intensity imaging via any of the three following approaches: SARA, HyperSARA, and Faceted HyperSARA. The Faceted HyperSARA library~\citep{fhs} is accessible from the
\href{https://basp-group.github.io/Puri-Psi/}{Puri-Psi} webpage~\citep{puripsi}.

\subsection{Library structure} \label{ssec:matlab:library}

The proposed library is composed of a core module, dubbed \texttt{Faceted-HyperSARA}, and two standalone sub-modules which are \texttt{RI-measurement-operator} and \texttt{SARA-dictionary}. 
Fig.~\ref{fig:diagram-fhs} provides an illustration of the structure of the library. Details of the three modules are provided in what follows.
\paragraph*{\texttt{RI-measurement-operator.}} The module covers all the functionalities associated with the formation of the measurement operator. These include the non-uniform Fourier transform, $w$-correction via $w$-projection, incorporation of available DDE solutions, and data dimensionality reduction via a visibility gridding-based scheme. The module is available at \url{https://basp-group.github.io/RI-measurement-operator}.

\paragraph*{\texttt{SARA-dictionary.}} The library provides faceted and non-faceted implementations of the SARA dictionary and comes with a detailed documentation, selected tests, and code examples. The module is available at \url{https://basp-group.github.io/SARA-dictionary}.

\paragraph*{\texttt{Faceted-HyperSARA.}} The core module, illustrated in Fig.~\ref{fig:pipeline-fhs}, enables the following procedures.
\begin{enumerate}
\item \emph{Data extraction from a measurement set}. A dedicated Python script enables the extraction of the RI data from one or multiple MS Tables into a collection of \texttt{.mat}. The script relies on \texttt{Pyxis}, standing for Python Extensions for Interferometry Scripting, which is part of the \href{http://meqtrees.net/}{MeqTrees software package}~\citep{Noordam10}. The extracted data files are then systematically fed as input to our MATLAB imaging pipeline; 

\item \emph{Imaging.} The library provides the implementation of the solvers (\emph{i.e.} algorithmic structures) underpinning the SARA, HyperSARA and Faceted HyperSARA approaches. Compared to~\citep{Thouvenin2021}, the solvers can seamlessly be applied to the inverse problem resulting from data dimensionality reduction via visibility gridding. The user interacts with the solvers solely through the dedicated imaging script \texttt{main\_input\_imaging.m}. The script requires a few user-defined parameters, described in Section~\ref{ssec:matlab:parameters} and calls the imaging function \texttt{imaging.m} which uses reliable default algorithmic parameters whenever left unspecified by the user. Finally, the library systematically uses the heuristic regularization parameters described in~\citep[Section 2]{Thouvenin2021} for the SARA, HyperSARA, and Faceted HyperSARA approaches, respectively.  
\end{enumerate} 
Associated documentation is available online \citep{fhs}.

\input{fig_library}

\subsection{Input parameters} \label{ssec:matlab:parameters}

The imaging pipeline depends on a number of parameters, which can be classified into the three following groups, depending on the level of interaction expected from the user.

\paragraph*{User-defined parameters.}
In order to run an image reconstruction task, a few parameters need to be specified in the \texttt{main\_input\_imaging.m} from the \texttt{Faceted-HyperSARA} module. These are specifically related to (i) the input data (\emph{e.g.} the name of target source, as used in the data extraction step, and the indices of the frequency channels to image), (ii) the output image cube (\emph{e.g.} the image size and the pixel size), (iii) the definition of the optimization problem~\eqref{eq:problem} (\emph{e.g.}, the selected approach among SARA, HyperSARA, and Faceted HyperSARA, and the scale factors to be applied to the heuristic values of the associated regularization parameters if different from 1), and (iv) algorithmic parameters (\emph{e.g.} the number of facets with the associated overlap for Faceted HyperSARA). A full list of these parameters is provided in the online documentation~\citep{fhs}.

\paragraph*{Default internal parameters.}
Several parameters, internal to the proposed imaging approach, are set to reliable default values in the auxiliary file \texttt{default\_parameters.json}. These include the kernel type and size taken for the NUFFT, and all the algorithmic parameters necessary to configure the reweighting procedure and inner PDFB algorithm~(see \citep[Algorithms 1 and 3]{Thouvenin2021}). A complete list of these parameters with the associated description is provided in the associated
documentation~\citep{fhs}.
%
Users are invited to rely on the default parameter configuration, unless specifically aware of their impact on the image reconstruction task. The default values can however be overridden by creating a new \texttt{.json} file, passed to the imaging script. In this case, users are strongly encouraged to start from the default configuration file and edit the values to be modified.

\paragraph*{Optional parameters.}
The library can take advantage of the results of a preliminary calibration step, where the pre-estimated DDE solutions are incorporated into the measurement operator or, alternatively, the DIE solutions are applied directly to the data (see Section~\ref{ssec:newfeatures:dde}). Residual visibilities can be leveraged to obtain estimates of the effective noise level as described in Section~\ref{ssec:newfeatures:noise}. 
Finally, an estimate of the image can be used to initialize the intensity imaging algorithm, for acceleration purposes.

\section{Data description and evaluation metrics} \label{sec:data_and_metrics}
To validate the proposed approach and MATLAB library at scale, a large RI wideband dataset is considered. Prior to the wideband intensity imaging, a pre-processing step, consisting in the monochromatic joint calibration and imaging is performed to enhance the accuracy of the measurement operator model, thus enabling higher precision imaging. This section provides the details of the utilized dataset and the pre-processing step. Evaluation criteria considered to assess the compared approaches are presented.

\subsection{Data description} \label{ssec:data_and_metrics:data}

The real data considered in this paper are part of wideband VLA observations of the celebrated radio galaxy Cyg~A, acquired in the frequency range 2--\SI{18}{\giga\hertz} over the years 2015-2016 (courtesy of R. Perley). A number of $\nband = 480$ channels in the C band were used. The selected frequencies span the range $[\nu_1,\nu_{480}]=[3.979,8.019]$~GHz, characterized by a channel-width $\delta \nu=~$\SI{8}{\mega\hertz} and a total bandwidth of \SI{4.04}{\giga\hertz}. 
The observations' phase center is located at the position $\rm{RA}=19\rm{h}~59\rm{mn}~28.356\rm{s}$ ($J2000$) and $\rm{DEC}=+40^{\circ}~44\arcmin~2.07\arcsec$. The dataset was acquired at four instances, corresponding to the frequency ranges $[\nu_1,\nu_{256}]=[3.979,{6.019}]$~GHz and $[\nu_{257},\nu_{480}]=[5.979,{8.019}]$~GHz, and VLA configurations A and C. The wideband data cube consists of 30 spectral windows, each composed of 16 channels spanning 128~MHz. The dataset contains approximately $10^6$ complex visibilities per channel (about $8\times10^5$ and $2\times 10^5$ measurements for configurations A and C, respectively), processed as double precision complex numbers. The data were flagged in CASA \citep{McMullin} in two steps. Firstly, known spectral regions affected by satellite transmission were removed. Secondly, sporadic data affected by radio frequency interference were removed using a simple thresholding procedure. The data have then been self-calibrated on a spectral window-basis, by alternating between an imaging step, which consists in the formation of a single image from the full spectral window, and a calibration step.

\subsection{Data pre-processing: monochromatic joint calibration and imaging} \label{ssec:data_and_metrics:calibration}

To improve the accuracy of the modeled measurement operator, a joint calibration and imaging pre-processing step was conducted for each frequency channel, separately~\citep{Dabbech2021}. The approach, originally proposed by~\citet{Repetti2017}, consists in the alternate estimation of the unknown DDEs and the image of interest, where a spatio-temporal smoothness prior is adopted for the DDEs, and an average sparsity prior in addition to non-negativity is considered for the image~\citep{Repetti2017spie,Thouvenin2018b}. 

Ideally, a wideband joint calibration and imaging approach would have been the most appropriate. Although preliminary work in this direction has been recently proposed
\citep{Dabbech2019}, a fully parallelizable algorithm remains to be developed in this context. 
Following the discussion in Section~\ref{ssec:background:imaging}, the calibration solutions can be easily integrated into the forward model~\eqref{eq:model}. Furthermore, the resulting estimated model visibilities can be leveraged to estimate the noise statistics, and consequently, the heuristic regularization parameters introduced in~\citep[Section 2]{Thouvenin2021}, as well as estimates of the $\ell_2$ bounds defining the data-fidelity terms in \eqref{eq:problem}, taken as the $\ell_2$ norms of the residual visibilities.

\subsection{Evaluation metrics} \label{ssec:data_and_metrics:metrics}

Imaging quality is evaluated by visually inspecting the reconstructed images\footnote{All figures in this article are displayed using the color scheme cubehelix proposed in \citet{cubehelix}.}, in comparison with SARA \citep{Onose2017} and the state-of-the-art wideband CLEAN-based method implemented in WSClean -- a C++ imager for RI~\citep{Offringa2017}. Note that HyperSARA \citep{Abdulaziz2019} is not considered as a benchmark approach due to its prohibitive runtime. 

For both Faceted HyperSARA and SARA, the analysis is conducted using the estimated model cube $\mathbf{X}$ and the naturally-weighted residual image cube $\mathbf{R}$ whose columns, indexed by $l \in \{1, \dotsc, \nband \}$, are given by $\mathbf{r}_l = \eta_l {\bs{\Phi}}_l^\dagger ({\mathbf{y}}_l - {\bs{\Phi}}_l {\mathbf{x}}_l)$, with ${\mathbf{y}}_l$ the naturally-weighted RI measurements. Let $\bs\delta \in \mathbb{R}^N$ denote an image with value 1 at the phase center and zero elsewhere. The normalization factor $\eta_l$ is such that the per-channel point spread function (PSF), given by $ \eta_l\bs{\Phi}_l^\dagger\bs{\Phi}_l{\bs{\delta}}$, has a peak value equal to 1. For WSClean, optimal results 
are obtained with Briggs weighting \citep{Briggs1995}. In addition to the Briggs-weighted residual cube $\widetilde{\mathbf{R}} = (\widetilde{\mathbf{r}}_l)_{1\leq l \leq L}$, the restored image cube ${\mathbf{T}}= ({\mathbf{t}_l})_{1\le l \le L}$ is considered. Its columns are defined as ${\mathbf{t}}_l = {\mathbf{x}}_l \ast \mathbf{c}_l + \widetilde{\mathbf{r}}_l$, for each channel index $l$ with $\mathbf{x}_l$ being the estimated model image (consisting of the CLEAN components) and $\mathbf{c}_l$ is the CLEAN restoring beam (typically a Gaussian fitted to the primary lobe of the associated PSF). We also consider in our analysis the image cube defined as the difference between the restored cube and the residual cube, whose columns are the model images convolved with the associated restoring beams. As a quantitative measure of data fidelity, the standard deviation of the recovered residual images is reported. 

All the methods compared in this paper were run on compute nodes of Cirrus, one of the UK's Tier2 HPC services\footnote{\url{https://epsrc.ukri.org/research/facilities/hpc/tier2/}}. Cirrus is an SGI ICE XA system composed of 280 compute nodes, each with two 2.1~GHz, 18-core, Intel Xeon E5-2695 (Broadwell) series processors. The compute nodes have 256~GB of memory shared between the two processors. The system has a single Infiniband FDR network connecting nodes with a bandwidth of 54.5~GB/s.

Computing cost is assessed in terms of the following metrics
\begin{enumerate}
  \item runtime per iteration ($\runp$). For SARA and Faceted HyperSARA, this metric is reported per PDFB iteration per sub-cube;
  \item (active) CPU time ($\cpup$), omitting the communication time between workers. For SARA and Faceted HyperSARA, the metric is reported per PDFB iteration per sub-cube; 
  \item reconstruction runtime (maximum over all subcubes), assuming all the subcubes are reconstructed in parallel ($\run$). This metric includes the communication time required between the facet and data workers. Note that the subcubes are reconstructed independently in parallel, as no communication is required between these imaging tasks;
  \item total (active) CPU time used to reconstruct the full image cube ($\cpu$).
\end{enumerate}
Per iteration metrics are reported in terms of their average and associated standard deviation over all iterations of the algorithm. On a final note, the pre-processing step is not included in the computing time reported for the different approaches.

\section{Experiment 1: wideband imaging at full spectral resolution} \label{sec:exp_real}

In this section, we study the imaging performance and the computational cost of Faceted HyperSARA via the reconstruction of a 15~GB image cube of Cyg~A from 7.4~GB of VLA data. Leveraging both the spectral and spatial faceting functionalities, the algorithm is mapped on 1440 CPU cores on the high performance computing system Cirrus. The performance of the proposed approach is evaluated in comparison with the monochromatic imaging approach SARA~\citep{Onose2017}. In this experiment, DDE estimates obtained from the pre-processing step described in Section~\ref{ssec:data_and_metrics:calibration} are considered. 
Since these estimates cannot be directly transferred to the WSClean software, the wideband CLEAN-based method is not considered for this experiment.

\subsection{Experimental setting} \label{ssec:exp_real:settings}

Imaging is conducted over the field-of-view (FoV) $\Omega_1=2.56 \arcmin \times 1.536\arcmin$ at the spatial resolution given by the pixel size $\delta x = 0.06\arcsec$ (in both directions), leading to $N=1536 \times 2560$. Full spectral resolution is considered, \emph{i.e.}, $\nband = 480$ channels. Hence, the wideband imaged cube is of size $ 1536 \times 2560 \times 480$. The selected pixel size is such that the spatial bandwidth of the recovered signal is up to 1.75 times the instrumental resolution at the highest channel $\nu_\nband = 8.019$~GHz, and 3.53 times the instrumental resolution at the lowest channel $\nu_1 =3.979$~GHz. Considering the same FoV and spatial resolution, the monochromatic joint calibration and imaging pre-processing step was performed separately on each frequency channel, providing antenna based and time dependent DDE estimates of spatial dimension $5\times 5$ in the spatial Fourier domain~\citep{Dabbech2021}. For all $ (c,l) \in \{1, \dotsc , C \} \times \{1, \dotsc, \nband_c \}$, these estimates were incorporated in the de-gridding matrices $\mathbf{G}_{c,l}$ (see related discussion in Section~\ref{ssec:background:imaging}). From now on, $\bs{\Phi}_{c,l}$ refer to the resulting measurement operators.

In both imaging algorithms, $B=2$ data blocks per channel are considered. The blocks are associated with the VLA configurations A and C, each having different noise statistics. For each data block indexed by $(c,l,b) \in \{ 1, \dotsc , C \} \times \{ 1, \dotsc , \nband_c \} \times \{ 1, \dotsc , \nblock \}$, the noise level $\sigma_{c,l,b}$ is estimated as the standard deviation of the residual visibilities $\mathbf{y}_{c,l,b}- {\bs{\Phi}_{c,l,b}}\mathbf{x}^{(0)}_{c,l}$, where $\mathbf{x}^{(0)}_{c,l}$ is the model image obtained from the pre-processing step. 
The $\ell_2$ bounds defining the data-fidelity constraints in~\eqref{eq:problem-split-subcube} are set to $\varepsilon_{c,l,b} = \| \mathbf{y}_{c,l,b}- {\bs{\Phi}_{c,l,b}}\mathbf{x}^{(0)}_{c,l} \|_2$. More specifically to Faceted HyperSARA, $C=16$ channel-interleaved sub-problems {are considered} with $L_c=30$ for any $c \in \{1, \dotsc , C \}$, using $\nfacet=10\times 6$ facets along the spatial dimension, with an overlap of about $50\%$. The resulting number of spatio-spectral facets is $Q \times C=960$. 
Given the estimates of the noise level, the regularization parameters of both algorithms are set from the heuristics proposed in \citet{Thouvenin2021} as follows. In Faceted HyperSARA, the low-rankness regularization parameter is fixed to the associated proposed heuristic, and the joint sparsity regularization parameter is set to 3 times the associated heuristic for optimal dynamic range. The regularization parameter of SARA is set to 3 times the proposed heuristic for the same reason. This deviation from the heuristics proposed in \citet{Thouvenin2021} can be explained by the need to compensate for residual calibration errors affecting the accuracy of the measurement operator model. For both imaging algorithms, a maximum of $5$ reweighting steps is considered.
To reduce the runtime to convergence, one could initialize the algorithms with the wideband model cube $\mathbf{X}^{(0)} = (\mathbf{X}^{(0)}_c)_{1\le c \le C}$ obtained from the joint DDE calibration and imaging step. However, in this experiment, to avoid any bias in the reported computational cost, both imaging algorithms have been initialized with a zero wideband cube.

\subsection{Results and discussion} \label{ssec:exp_real:results}

\subsubsection{Imaging quality} \label{sssec:exp_real:results:quality}
\input{fig_cyga_exp1_ch102}
\input{fig_cyga_exp1_ch459}

To assess the imaging quality of Faceted HyperSARA in comparison with SARA,
we examine the estimated images of the selected frequency channels $\nu_{102} = 4.787$~GHz and $\nu_{459} = 7.851$~GHz, displayed
in Figures~\ref{fig:cyga102} and \ref{fig:cyga459} respectively. The channels correspond to the respective channel indices 7 and 29 of sub-problems 6 and 11. The model images, focusing on Cyg~A, are displayed in $\log_{10}$ scale, and overlaid with the corresponding residual images. Zooms on selected regions of the radio galaxy have been included to enhance the visual comparison.

Although both algorithms yield very comparable reconstructions for high-intensity pixels ($\ge 1$~mJy, see the hotspot regions), Faceted HyperSARA (top rows of Figures~\ref{fig:cyga102}-\ref{fig:cyga459}) supersedes SARA (bottom rows of Figures~\ref{fig:cyga102}-\ref{fig:cyga459}) at low intensity by capturing a more physically-consistent structure. For instance, the annulus at the high end of the western lobe is consistently recovered by Faceted HyperSARA at both frequency channels, while it is fully dominated by artefacts in the form of stripes in the SARA model image at the low frequency channel. This behaviour is also observed in other regions of the radio galaxy (\emph{e.g.} see both zooms of Figure~\ref{fig:cyga459}). Moreover, the improved dynamic range of Faceted HyperSARA can be observed in the recovery of faint emission ($\le 0.01$~mJy), such as additional segments of the jets and the faint extended emission at the ends of both lobes (see both zooms of Figure~\ref{fig:cyga102}). These emissions were previously detected in Cyg~A image at S band formed from highly sensitive monochromatic VLA observations \citep{Dabbech2021}. The superior imaging quality of Faceted HyperSARA is the result of the correlation imposed across channels, making it more robust to calibration errors than its monochromatic counterpart. Note that calibration artefacts observed in SARA images can be reduced by enforcing more sparsity through a higher regularization parameter, yet at the expense of a much smoother reconstruction.
Importantly, the spatial tessellation underlying Faceted HyperSARA image model does not introduce any discernible artefact in the estimated images.

Full FoV residual images obtained with the two algorithms are reported in linear scale overlaying the estimated model images in Figures~\ref{fig:cyga102}--\ref{fig:cyga459}. At the selected channels, both algorithms exhibit large artefacts in the form of a negative hole at the inner core of the galaxy, very likely induced by residual calibration errors. Interestingly, at the low frequency channel (Figure~\ref{fig:cyga102}), Faceted HyperSARA residual image exhibits stripe-like structure at both lobes of Cyg~A, similar to the artefacts observed in SARA model image in the same region. The fact that these artefacts are left out of Faceted HyperSARA model image confirms the robustness of its underpinning wideband image model to channel-based calibration errors. Finally, the standard deviations of the residual images obtained by the two algorithms, reported in the captions of Figures~\ref{fig:cyga102}-\ref{fig:cyga459}, are highly comparable, though nearly five times above the noise estimate proposed in \citet{Thouvenin2021}, at both frequency channels, due to calibration errors.

Full image cubes are made available online in FITS and GIF formats~\citep{faceting}. Their visual examination indicates a generally smooth spectral behaviour of Faceted HyperSARA wideband image, in line with expected power law decays associated with synchrotron emission. In contrast, SARA wideband image exhibits prominent channel-varying fluctuations mostly in the form of stripe-like artefacts, confirming the above analysis. Although a spectral index map can be directly inferred from the reconstructed images, such analysis is beyond the scope of this work. Nonetheless, we provide a spectral analysis based on selected emissions from different regions in the radio galaxy.
For Faceted HyperSARA, the spectra at different pixel positions are generally consistent with a continuum emission following a typical power-law decay, with the exception of a noticeable spectral discontinuity observed at channel $\nu_{257}=5.979$~GHz. Since the data were acquired separately in the frequency ranges $[\nu_1,\nu_{256}]=[3.979,{6.019}]$~GHz and $[\nu_{257},\nu_\nband]=[5.979,{8.019}]$~GHz, this discontinuity is most likely due to differences in the calibration errors and noise statistics between the two frequency ranges. A small-scale spectral discontinuity demarcating the spectral windows can be further noticed, which most likely results from calibration errors induced by the initial calibration procedure conducted on a spectral window basis (see Section~\ref{ssec:data_and_metrics:data}). As for SARA, additional irregularities of the spectra at different pixel positions are observed, particularly important at pixel positions with intensity values below 1~mJy, reflecting the channel-varying fluctuations due to the inherent calibration errors and the lack of correlation across the channels.

\subsubsection{Computing cost} \label{sssec:exp_real:results:cost}

\input{table_timing_cyga.tex}
Faceted HyperSARA and SARA were run on multiple nodes of Cirrus. For Faceted HyperSARA, each sub-problem~\eqref{eq:problem-split-subcube} (composed of 30 channels) was solved using 91 CPU cores, distributed as follows; 1 master CPU core, 30 CPU cores to process the $2\times 30$ data-fidelity terms (four data-fidelity blocks per core), and 60 CPU cores to handle the 60 spatio-spectral facets. As for SARA, the method was applied in the parallel configuration proposed by~\citet{Onose2016}. More precisely, each channel was imaged using 12 CPU cores, nine of which were exploited by the average sparsity terms (associated with the nine bases of the SARA dictionary).
The computing time and resources required by Faceted HyperSARA and SARA are summarized in Table~\ref{tab:cyga}. Faceted HyperSARA required nearly $50\%$ more resources than SARA in terms of $\cpu$ hours in this case. This could be explained by the larger number of workers used for the Faceted HyperSARA regularization and the higher number of iterations required by the algorithm to reach convergence, as already observed in~\cite{Thouvenin2021}.

\section{Experiment 2: wideband imaging with dimensionality reduction} \label{sec:exp_DR}

In the previous section, spectral faceting has been leveraged to divide the imaging problem into smaller, independent sub-problems. Faceting allowed a wideband model cube of Cyg~A to be formed at the full spectral resolution of the data. The image cube estimate exhibits a relatively flat spectrum across the frequency range of interest. This observation suggests that a much coarser spectral resolution could have been adopted \emph{a priori} without violating the Nyquist limit. Data points from consecutive channels can therefore be combined to form an image cube with a lower spectral dimension. In this setting, the amount of data per \textit{effective} channel increases significantly, leading to large memory requirements to reconstruct the image cube. Memory costs required to store the {de-gridding} matrix $\mathbf{G}$ involved in the measurement operator (up to 2.5~TB for the 7.4~GB data utilized in Section~\ref{sec:exp_real}) are particularly large. To tackle this challenge, we leverage the data dimensionality reduction procedure relying on a visibility gridding-based scheme described in Section~\ref{ssec:newfeatures:gridding}. The resulting joint image and data dimensionality reduction can significantly reduce both memory requirements and computing cost to form RI image cubes. In this section, we study the performance of Faceted HyperSARA in comparison with the CLEAN-based wideband imaging algorithm in WSClean.

\subsection{Experimental setting} \label{ssec:exp_DR:setting}

Given the reconstruction results reported in Section~\ref{ssec:exp_real:results} at the spectral resolution $\delta \nu =8$~MHz, we consider herein a coarser spectral sampling at the resolution $\widehat{\delta \nu}=128$~MHz. In other words, the 16 consecutive channels composing each spectral window are combined to form a single effective channel image. In this setting, a joint DIE calibration and imaging pre-processing step was applied to each of the effective channels separately, with the aim to improve the accuracy of the measurement operator while ensuring a fair comparison with the state-of-the-art CLEAN-based wideband approach in WSClean. As in the previous experiment, the FoV adopted in the pre-processing step is $\Omega_1=2.56 \arcmin \times 1.536\arcmin$, with the spatial resolution given by the pixel size $\delta x = 0.06\arcsec$ (in both directions). Antenna gain solutions, which are now reduced to complex values, were applied directly to the data. The \textit{corrected} data were then easily stored in the measurement sets and thus leveraged by both Faceted HyperSARA and the wideband imaging approach in WSClean. Note that the $29^{\textrm{th}}$ spectral window (spanning the frequency range $[7.767,7.895]$ GHz) was removed due to large calibration errors. The target wideband cube is therefore composed of $\widehat{\nband}=29$ effective channels. Despite the absence of DDE correction, this pre-processing step results in a good image reconstruction quality as will be shown in the following paragraphs.

For wideband imaging, we consider the FoV $ \Omega_2= 2.1709 \arcmin \times 3.6181\arcmin$, with a spatial resolution given by the pixel size $\delta x=0.0424\arcsec$ (in both directions). The spatial bandwidth of the recovered signal is up to {$2.4906$} times the instrumental resolution at the highest effective channel $\widehat{\nu}_{29}=7.959$~GHz, and {$4.9245$} times the instrumental resolution at the lowest effective channel $\widehat{\nu}_1=4.039$~GHz. This wideband imaging setting is different from the pre-processing step where a coarser resolution and smaller FoV are considered to reduce its computational cost. The target wideband image cube is of size $N \times \widehat{\nband}=3072 \times 5120 \times 29$. Given the significant reduction in the spectral dimension, only one spectral facet is considered in Faceted HyperSARA (\emph{i.e.} $C =1$, and $\nband_C =\widehat{\nband}=29$). The index $c$ will thus be omitted in the following paragraphs for simplicity.

Initially composed of about $4.5 \times 10^8$ points, the data are reduced down to $6.5\times 10^{7}$ points after gridding, that is about 7-fold reduction in data volume, from 7.2~GB down to 1.04~GB. Since DIE solutions are adopted in this experiment, the memory required to store the de-gridding matrices $(\mathbf{G}_{l,b})_{{ (1,1) \leq (l,b) \leq (\widehat{\nband}, \nblock)}}$ is about 530~GB, a decrease of a factor $\approx$ 4.55 in comparison with the experiment conducted in Section~\ref{sec:exp_real}. 
The data dimensionality reduction yields a decrease in the memory requirements of the resulting measurement operator, with the underlying holographic matrices $(\bs{\Lambda}_{l,b}\mathbf{H}_{l,b})_{{ (1,1) \leq (l,b) \leq (\widehat{\nband}, \nblock)}}$ occupying about 108~GB, a 4.8-fold reduction of the memory needed to store the de-gridding matrices.  

For each effective channel, the data associated with the two VLA configurations and the different channels forming a spectral window are aggregated into $\nblock=1$ block after gridding. The noise level after gridding and the associated data constraints are set from the model visibilities $\bar{\mathbf{y}}_{l,b}$, obtained from the pre-processing step. More precisely, for each data block indexed by $\left(l,b\right)\in \{ 1, \dotsc , \widehat{\nband} \} \times \{ 1, \dotsc , \nblock \}$, the associated $\ell_2$ bound is estimated as
$\varepsilon^{\prime}_{l,b} = \| \mathbf{y}^{\prime}_{l,b}- {\bs{\Lambda}_{l,b}\mathbf{G}_{l,b}^\dagger \bs{\Theta}^\dagger_{l,b}} \bar{\mathbf{y}}_{l,b} \|_2 $. 

Faceted HyperSARA was applied with $\nfacet = 10\times 6$,~\emph{i.e.} $Q\times C=60$ spatio-spectral facets with an overlap of about $50\%$. Using the estimates of the noise level, the regularization parameters are fixed using the heuristics proposed in \citet{Thouvenin2021}. The low-rankness regularization parameter is set to the proposed heuristic, whereas the joint sparsity regularization parameter is set to 0.1 times its associated heuristic for optimal resolution. Once again, the choice of the regularization parameter associated with the average joint sparsity prior deviates from the proposed heuristic, which can be explained by the inaccuracy of the measurement operator in the presence of residual calibration errors. In the re-weighting procedure, a maximum of $5$ reweighting steps was considered. Finally, Faceted HyperSARA was initialized with a zero wideband cube, since the monochromatic joint DIE calibration and imaging pre-processing step was conducted at a coarser resolution and over a smaller FoV compared to the wideband imaging step. 

For WSClean, Briggs weighting with the robust parameter set to $-0.25$ was considered to enhance the overall resolution. The wideband deconvolution was performed using a spectral polynomial fit of order 3 and $10^6$ maximum number of iterations. The command used in WSClean is reported in Appendix~\ref{appendix:wsclean}.

\subsection{Results and discussion} \label{ssec:exp_DR:results}

\subsubsection{Imaging quality} \label{sssec:exp_DR:results:quality}
\input{fig_cyga_exp2_ch10.tex}
\input{fig_cyga_exp2_ch29.tex}

Reconstructed images at two selected channels $\widehat{\nu}_{10}=5.191$~GHz and $\widehat{\nu}_{29}=7.959$~GHz are reported in Figures~\ref{fig:cyga_dr_ch10} and~\ref{fig:cyga_dr_ch29} respectively. From top to bottom, are shown the estimated model image of Faceted HyperSARA, the restored image of the CLEAN-based approach in WSClean and the associated model image convolved with a restoring beam and obtained by subtracting the residual image from the restored image. All the images are displayed in $\log_{10}$ scale. WSClean images are normalized by the flux of their associated restoring beam, so that all images are in the same unit Jy/pixel. In addition, the negative pixels in WSClean images are set to zero for visualization purposes. All images are displayed with zooms on selected regions of the radio galaxy to better assess their dynamic range and resolution. Finally, full FoV residual images, displayed in linear scale, are overlaying the respective estimated model and restored images of Faceted HyperSARA and WSClean.

Looking at the estimated model images of Faceted HyperSARA (top rows of Figures~\ref{fig:cyga_dr_ch10}-\ref{fig:cyga_dr_ch29}) in comparison with the restored images of WSClean (middle rows of Figures~\ref{fig:cyga_dr_ch10}-\ref{fig:cyga_dr_ch29}), one can clearly see the superior performance of the proposed imaging algorithm, in terms of both resolution (see the zoomed region from the eastern hotspot, displayed on the bottom right of each panel of Figure~\ref{fig:cyga_dr_ch10}) and dynamic range, particularly noticeable at the tails of the lobes (see the zoomed regions of the tail of the western lobe and the jet emission, displayed on the top right of each panel of Figure~\ref{fig:cyga_dr_ch29}). By design, the dynamic range of the WSClean restored image is severely limited by the addition of the residual image, a standard post-processing step for CLEAN-based approaches. Indeed, when the latter is removed, the model image convolved with a restoring beam (bottom rows of Figures~\ref{fig:cyga_dr_ch10}-\ref{fig:cyga_dr_ch29}) reveals the captured faint and extended emission. Nonetheless, the image exhibits a large number of artefacts in the form of multi-resolution 2D Gaussian-like components, not to mention its physically-inconsistent negative components.\footnote{Further tuning WSClean by increasing the auto-masking threshold has shown to mitigate the number of the artefacts in the model image convolved with the restoring beam, at the expense of resolution, particularly noticeable at the jets and the ends of both lobes. This however has no impact on the quality of the restored image, dominated by the calibration artefacts left in the residual image.}
Calibration errors can be observed in the reconstructions of both algorithms. Firstly, these artefacts take the form of structures emanating from the hotspots in the model images of Faceted HyperSARA, and ringing artefacts around the hotspots in WSClean images. Similar structures were observed in reconstructions of Cyg~A at similar frequencies~\citep{Dabbech2021} and were attributed to pointing errors. Secondly, structures around the core of the radio galaxy are observed in the low frequency-channel image recovered by Faceted HyperSARA (top row of Figure~\ref{fig:cyga_dr_ch10}) and WSClean reconstructed images at both channels (bottom rows of Figures~\ref{fig:cyga_dr_ch10}-\ref{fig:cyga_dr_ch29}).

Residual images displayed in linear scale, obtained with both algorithms, are displayed over the associated images (top and middle rows of each figure). Note that the images associated with both methods are not obtained using the same measurement operator. Faceted HyperSARA adopts a naturally weighted de-gridding operator. In contrast, the CLEAN-based approach adopts a Briggs-weighted de-gridding measurement operator, thus emphasizing the high spatial frequency content. The resulting residual images are therefore not comparable in terms of patterns and structure. On the one hand, ringing structures centered at the hotspots are noticed in Faceted HyperSARA residual images for the two selected channels. On the other hand, the residual images of WSClean present high signal at the pixel positions of the hotspots. One can also notice ringing structures of somewhat smaller scale than those observed in Faceted HyperSARA residual images due to the different data weighting schemes adopted in both algorithms. The standard deviations of both residual images, computed after normalization with the flux of the associated restoring beams and reported in the captions of Figures~\ref{fig:cyga_dr_ch10}-\ref{fig:cyga_dr_ch29}, indicate the higher fidelity to data of Faceted HyperSARA.

Examining the reconstructed image cubes, available online in both FITs and GIF formats~\citep{faceting}, confirms the analysis reported above.
As a single effective channel per spectral window is considered, the different emissions in Cyg~A recovered by our proposed approach exhibit smooth spectra in general, confirming a typical power-law decay. One spectral discontinuity is observed the effective channel $\widehat{\nu}_{17}=6.039$~GHz, that is at the intersection between the two effective frequency ranges $[\widehat{\nu_1}, \widehat{\nu}_{16}]= [4,039, 5.959]$ GHz and $[\widehat{\nu}_{17},\widehat{\nu}_{29}]=[6.039, 7.959]$ GHz. As already reported in Section~\ref{sssec:exp_real:results:quality}, this behaviour is most likely induced by the different calibration errors affecting the independently acquired observations in these two frequency ranges. On a final note, the examination of the spectra at different pixel positions in WSClean restored cube is hampered by the channel-varying errors present in the associated residual cube.

\subsubsection{Computing cost} \label{sssec:exp_DR:results:cost}

\input{table_timing_cyga_dr.tex}

Faceted HyperSARA exploited 90 CPU cores for about 32 hours as follows; 1 master CPU core, 60 CPU cores handling the spatial facets and 29 CPU cores handling the data-fidelity terms, and 2903 $\cpu$ hours in total. Further details on the computational and memory cost are reported in Table~\ref{tab:cyga_dr}. 
WSClean was run with 29 CPU cores for about 17 hours, yielding a total computing cost of about 493 $\cpu$ hours (assuming all the cores are active the whole time). Though the numbers are sensitive to the resources' allocation, the total $\cpu$ time of the C++ RI imager remains significantly lower than that required by the MATLAB implementation of Faceted HyperSARA.

\section{Conclusion and future work} \label{sec:conclusion}

In this paper, new functionality of Faceted HyperSARA, a wideband intensity imaging approach for RI proposed in the companion paper~\citep{Thouvenin2021}, has been introduced for enhanced robustness and scalability when accommodating large scale RI data. On the one hand, available direction-dependent models, including the $w$-terms, are supported in the measurement operator to enhance its accuracy. In addition, an adaptive strategy for on the fly estimation of the effective noise level can be activated, in case of unknown noise levels due to the imperfect calibration, ensuring the robustness of the imaging algorithm in terms of reconstruction quality. On the other hand, significant reduction of the dimension of the data is achieved via a visibility gridding-based scheme.

A fully documented MATLAB library (available on {\textsc{Github}} \url{https://basp-group.github.io/Faceted-HyperSARA/}~\citep{fhs}) was introduced as part of a wider effort towards a larger adoption of the algorithms of the SARA family by the community.

The enhanced algorithm and code are assessed on two real data experiments in comparison with SARA~\citep{Carrillo2012,Onose2017} and the wideband CLEAN-based method implemented in WSClean~\citep{Offringa2017}. Firstly, leveraging the power of a large scale High Performance Computing system, Faceted HyperSARA was validated on the reconstruction of a 15~GB image cube of Cyg~A from 7.4~GB of VLA data. The obtained results are a practical proof of concept of the parallelization flexibility of Faceted HyperSARA, which is also shown to provide a significant improvement in the imaging quality with respect to SARA. Secondly, the image formation process of Cyg~A was revisited considering an important reduction of the image dimension via a reduction in the spectral resolution, and of the data dimension via a visibility gridding-based scheme. In this setting, significant reductions in memory requirements and computing cost are observed. More interestingly, Faceted HyperSARA has shown to be effective in recovering a higher quality wideband image, compared with the state-of-the-art CLEAN-based approach of the WSClean software.

Having addressed the computational bottlenecks raised by both the volume of the data and the size of the image, future work could contemplate the use of a faceted Fourier transform to improve the data and image locality in the proposed algorithm~\citep{Wortmann2021}. Other perspectives include the extension of the approach to polarization imaging and the development of a production C++ version of Faceted HyperSARA, building on the existing C++ version of HyperSARA (see the \href{https://basp-group.github.io/Puri-Psi/}{Puri-Psi} webpage~\citep{puripsi}). Ultimately, beyond image estimation, the faceted approach should also be integrated into a joint calibration and imaging approach~\citep{Repetti2017} and combined with the recently proposed methods for uncertainty quantification by convex optimization~\citep[\emph{e.g.}][]{Repetti2018,Repetti2019,Abdulaziz2019b}.

\section*{Acknowledgements}

The first two authors contributed equally to this work. The authors warmly thank R.A.~Perley (NRAO, USA) for providing the VLA observations of Cyg~A. The National Radio Astronomy Observatory is a facility of the National Science Foundation operated under cooperative agreement by Associated Universities, Inc. We also thank H.L.~Bester (SARAO, South Africa) for useful discussions on data dimensionality reduction and A.~Wilber (Heriot-Watt University, UK) for helpful feedback on the MATLAB library. This work was supported by EPSRC, grants EP/M011089/1, EP/M008843/1, EP/M019306/1, EP/T028270/1, EP/T028351/1, and ST/W000970/1 the Swiss-South Africa Joint Research Program (IZLSZ2\_170863/1), and used the Cirrus UK National Tier-2 HPC Service at EPCC (\url{http://www.cirrus.ac.uk}) funded by the University of Edinburgh and EPSRC (EP/P020267/1).

\section{Data availability}

The images underlying this article are available from the Heriot-Watt University research portal, with the digital object identifier \texttt{\href{https://researchportal.hw.ac.uk/en/datasets/faceted-hypersara-for-ri-wideband-intensity-imaging-cygnus-a-wide}{10.17861/f127fd8d-9394-4d74-abf0-7af35bd7192e}}. MATLAB codes associated with this article are made available online on the Faceted HyperSARA page \url{https://github.com/basp-group/Faceted-HyperSARA} of the Puri-Psi library~\cite{puripsi}.


\bibliographystyle{mnras}
\bibliography{strings_all_ref,all_ref}

\appendix

\section{WSClean command used in experiment 2} \label{appendix:wsclean}

The command used in WSClean to obtain the results of Experiment 2, displayed in Figures~\ref{fig:cyga_dr_ch10} and ~\ref{fig:cyga_dr_ch29}, is provided below.
\begin{minted}{bash}
    wsclean -join-channels -channels-out 29 \
    -spws 0,1,2,3,4,5,6,7,8,9,10,11,12,13,14,\
    15,16,17,18,19,20,21,22,23,24,25,26,27,29 \ 
    -fit-spectral-pol 3 -reorder -scale 0.0424arcsec \ 
    -weight briggs -0.25 -data-column DATA \ 
    -pol I -size 5120 3072 -niter 1000000 \
    -nwlayers 1 -field 2 -mgain 0.8 -gain 0.08 \
    -auto-mask 3 -auto-threshold 1 \
    -no-small-inversion -multiscale \
    -name CYGA-EXP2 CYGA.MS
\end{minted}


\bsp	
\label{lastpage}
\end{document}

%% file: tab_notations.tex
\begin{table}
    \centering
    \caption{Problem dimensions and mathematical notations}
    \label{tab:notations}
    \label{tab:maths_notations}
    \begin{tabular}{ll} \toprule
        \multicolumn{2}{l}{\emph{Problem dimensions}} \\ \midrule
        $\nbpix$ & image size \\
        $\nfourier$ & size of the 2D oversampled Fourier space \\
        $\nblock$ & number of data blocks per channel \\
        $\nband$ & number of spectral channels \\
        $\ndata$ & number of visibilities \\
        $\ncube$ & number of spectral sub-cubes \\
        $\nfacet$ & number of spatial facets \\ \midrule
        \multicolumn{2}{l}{\emph{Mathematical notations}} \\ \midrule
        $\mathbf{Z}^\dagger$ & adjoint (conjugate transpose) of the matrix $\mathbf{Z}$ \\
        $\iota_C$ & indicator function of the set $C$ \\
        \bottomrule
    \end{tabular}
\end{table}

%% file: fig_library.tex
\begin{figure*}
    \centering 
    \begin{subfigure}[b]{0.89\textwidth}
        \centering
        \includegraphics[width=\textwidth,keepaspectratio]{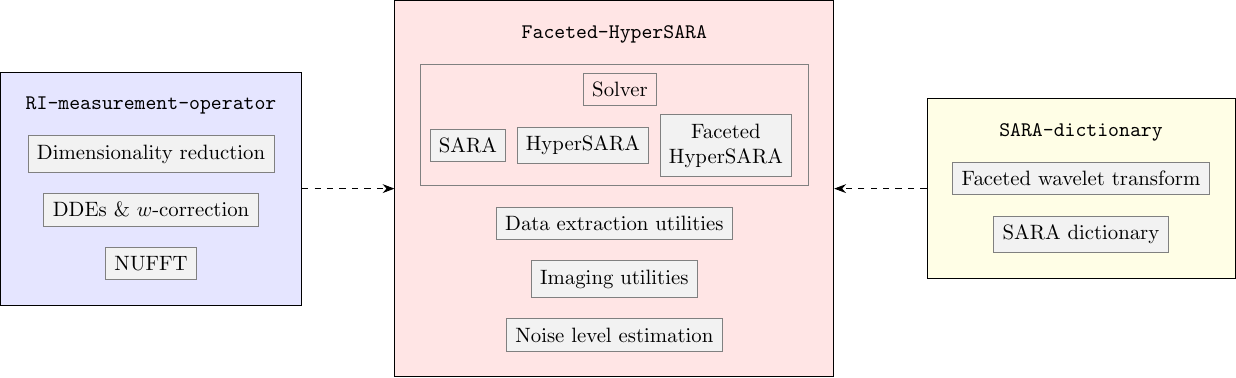}
        \caption{Library structure \vspace{1cm}}
        \label{fig:diagram-fhs}
    \end{subfigure}
    \begin{subfigure}[b]{0.99\textwidth}
        \centering
        \includegraphics[width=\textwidth,keepaspectratio]{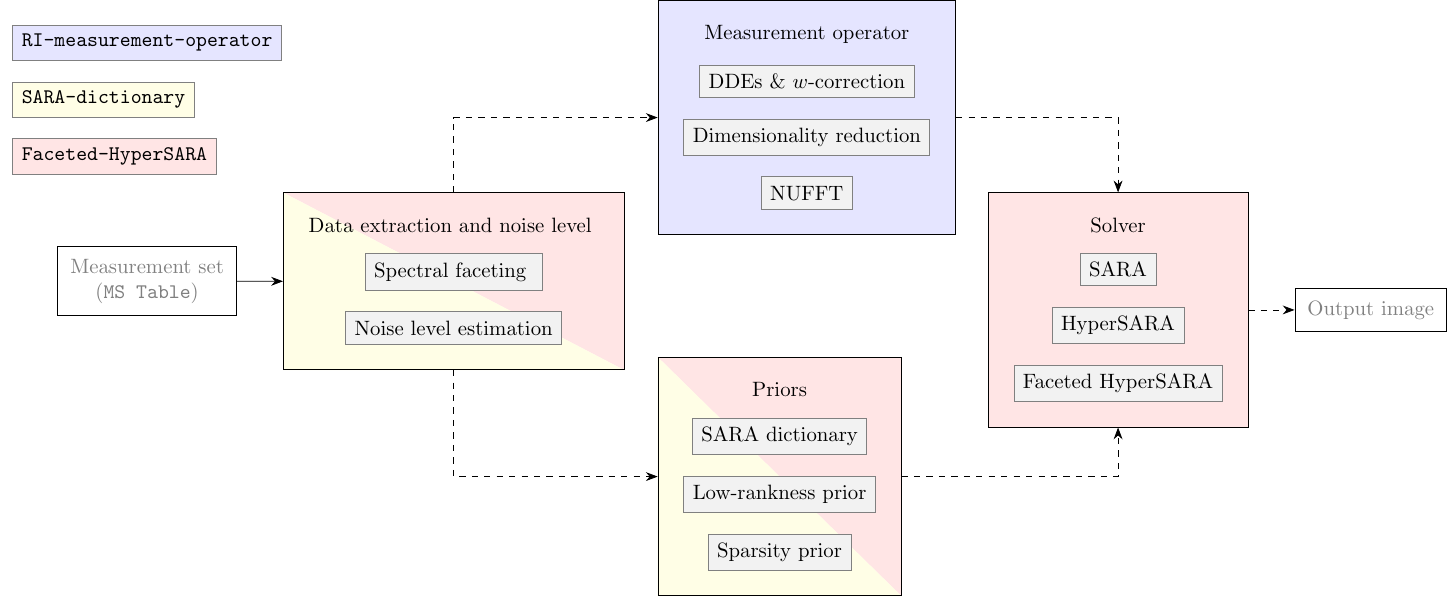}
        \caption{Imaging pipeline}
        \label{fig:pipeline-fhs}
    \end{subfigure}
    \caption{Proposed library structure and illustration of the Faceted HyperSARA imaging pipeline. The core module (\texttt{Faceted-Hypersara}) appears in red, whereas auxiliary submodules (\texttt{RI-measurement-operator} and \texttt{SARA-dictionary}) appear in blue and yellow, respectively.}
    \label{fig:library}
\end{figure*}

%% file: fig_cyga_exp1_ch102.tex
\begin{figure*}
  \centering
  \includegraphics[width=0.64\textheight,keepaspectratio,trim=0cm 1.6cm 0cm 0.8cm, clip]{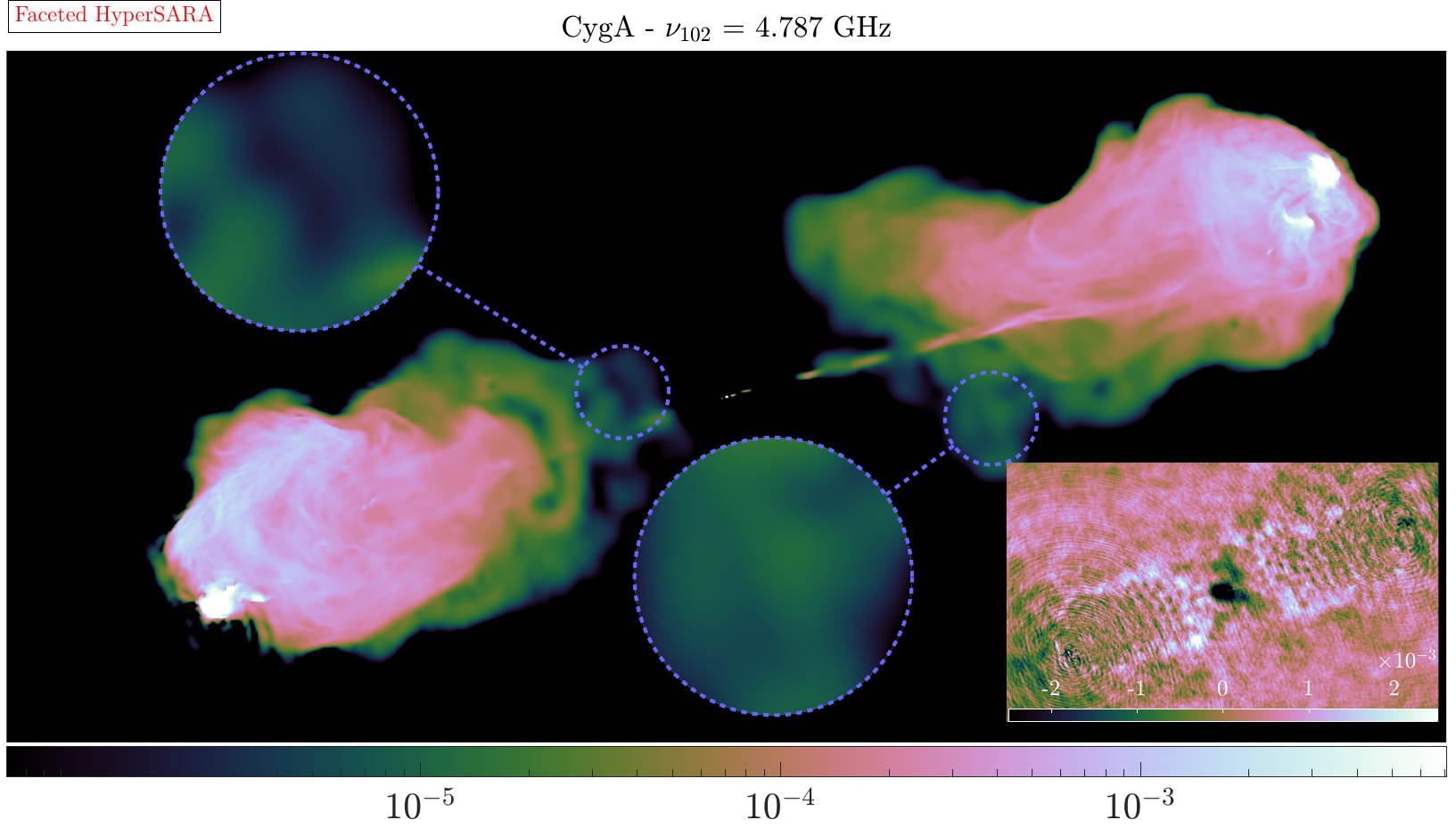}
  \includegraphics[width=0.64\textheight,keepaspectratio,trim=0cm 0cm 0cm 0.8cm, clip]{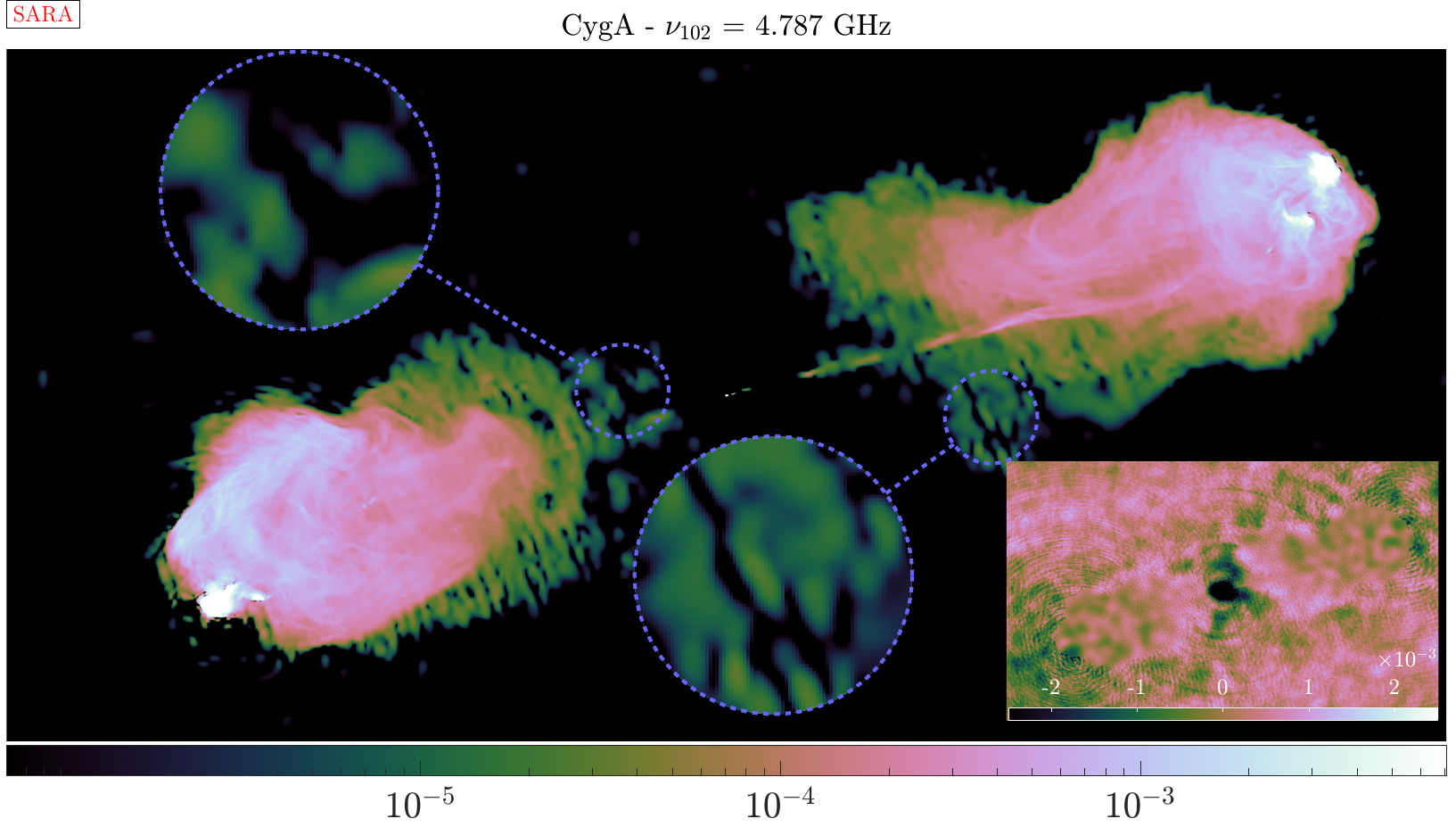}

  \caption{Experiment 1: Imaging results of Cyg A at the spectral resolution $\delta \nu=8$ MHz and spatial resolution $\delta x=0.06\arcsec$. Reconstructions of channel $\nu_{102} = 4.787$ GHz. Top and bottom rows: estimated model images of the proposed Faceted HyperSARA with ($\nfacet=60$, $C=16$) and SARA, respectively, in units of Jy/pixel and displayed in $\log_{10}$ scale. They are overlaid with zooms on several regions of the radio galaxy, highlighted in dashed circles to enhance the visual comparison between the two algorithms, and the associated residual images displayed in linear scale. The respective standard deviations of Faceted HyperSARA and SARA, after a normalization by the flux of the restoring beam, are $ 0.021$~mJy/pixel and $0.020$~mJy/pixel. Full image cubes are available online~\citep{faceting}.}

  \label{fig:cyga102}
\end{figure*}

%% file: fig_cyga_exp1_ch459.tex
\begin{figure*}
  \centering
  \includegraphics[width=0.64\textheight,keepaspectratio,trim=0cm 1.6cm 0cm 0.8cm, clip]{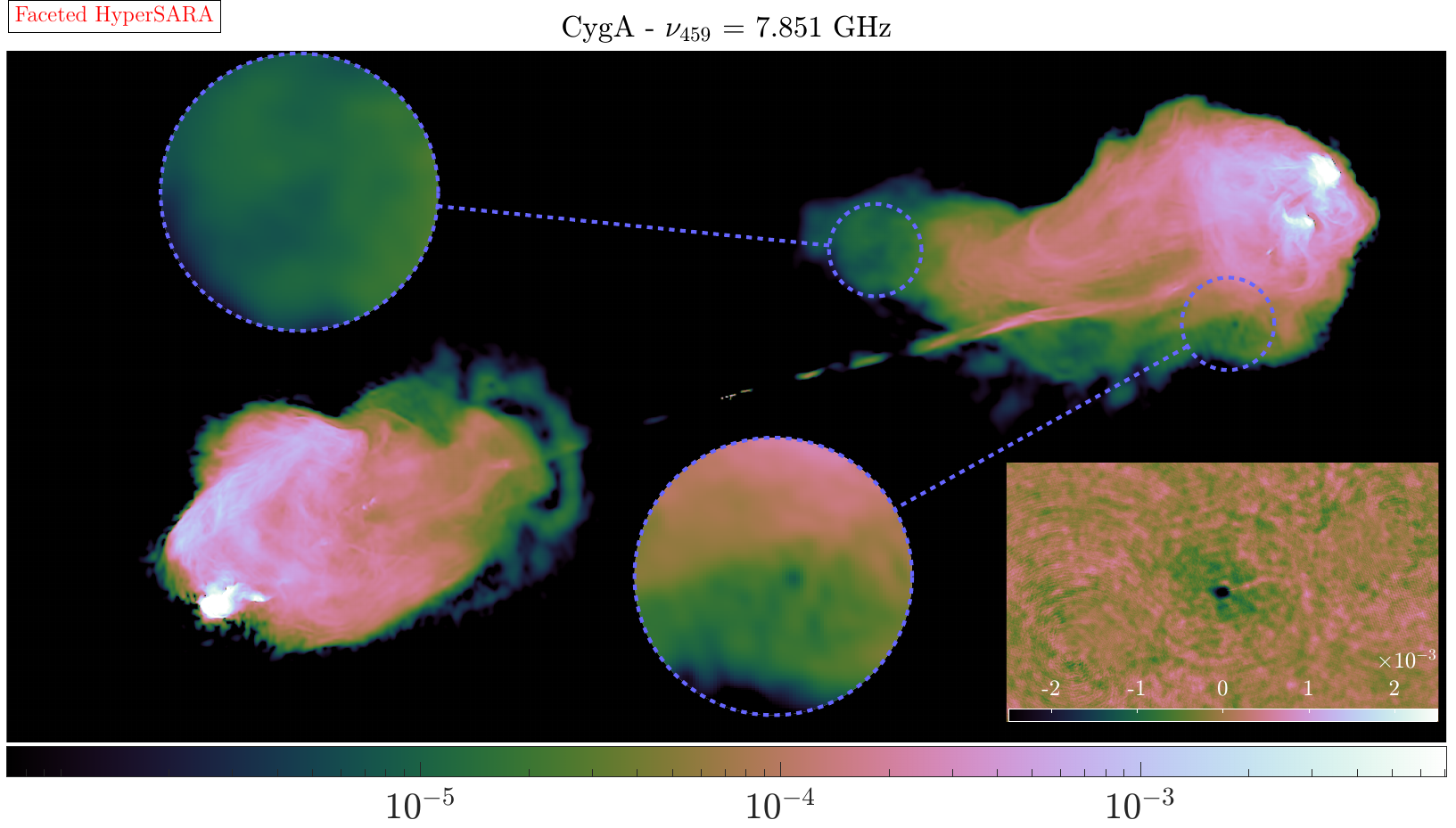}
  \includegraphics[width=0.64\textheight,keepaspectratio,trim=0cm 0cm 0cm 0.8cm, clip]{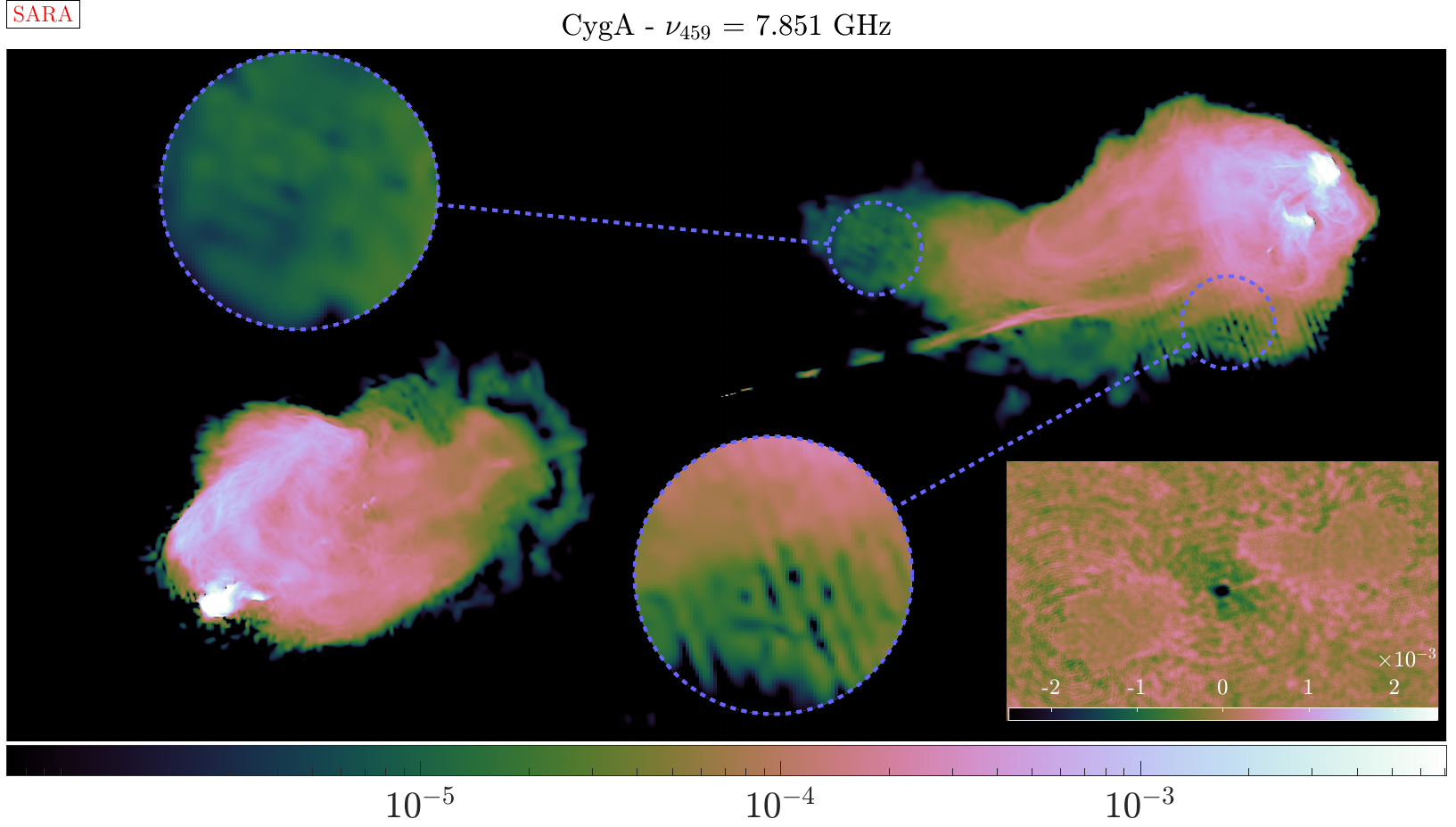}

  \caption{Experiment 1: Imaging results of Cyg A at the spectral resolution $\delta \nu=8$ MHz and spatial resolution $\delta x=0.06\arcsec$. Reconstructions of channel $\nu_{459} = 7.851$ GHz. Top and bottom rows: estimated model images of the proposed Faceted HyperSARA  with ($\nfacet=60$, $C=16$) and SARA, respectively, in units of Jy/pixel and displayed in $\log_{10}$ scale. They are overlaid with zooms on several regions of the radio galaxy, highlighted in dashed circles to enhance the visual comparison between the two algorithms, and the associated residual images displayed in linear scale. The respective standard deviations of Faceted HyperSARA and SARA, after a normalization by the flux of the restoring beam, are $0.026$~mJy/pixel and $0.027$~mJy/pixel. Full image cubes are available online~\citep{faceting}.}
  \label{fig:cyga459}
\end{figure*}

%% file: table_timing_cyga.tex
\begin{table}
    \centering
    
    \begin{tabular}{lll} \toprule
        &Faceted HyperSARA & SARA \\ \midrule
        CPU cores ($N_c$)    & 91 & 12 \\ 
        Total CPU cores (=$N_{c}\times C$)  & 1456 & 5760 \\
        average PDFB iterations  & 4598 & 2513 \\
        $\runp$ (s) &   20.05 ($\pm 0.81$) & 6.25 ($\pm 0.56$)  \\ 
        $\run$ (h)  &  34.01
        & 8.04 
        \\
        $\cpup$ (s) & 1824.55 ($\pm 73.71$) & 75.02 ($\pm 6.74$)  \\ 
        $\cpu$ (h) &  37324  & 25141 \\ %
        Memory allocated to compute &  \multirow{2}{*}{3.36 TB} & \multirow{2}{*}{3.36 TB} \\
        the $\mathbf{G}_{c,l,b}$ matrices & & \\
        Memory occupied by &  \multirow{2}{*}{2.5 TB} & \multirow{2}{*}{2.5 TB} \\
        the $\mathbf{G}_{c,l,b}$ matrices & & \\
        \bottomrule
    \end{tabular}
    \caption{Experiment 1: Computing cost of Cyg A imaging at the spectral resolution $\delta \nu=8$ MHz and spatial resolution $\delta x=0.06\arcsec$. Results are reported for Faceted HyperSARA (with $C=16$, $Q=60$) and SARA (\emph{i.e.} with $C=L$ and $Q=1$ ) in terms of: number of CPU cores, average number of PDFB iterations (across the $\ncube$ independent problems), runtime per iteration per sub-cube and total runtime ($\runp$ in seconds with the associated standard deviation, $\run$ in hours), CPU time per iteration per sub-cube and total CPU time ($\cpup$ in seconds with the associated standard deviation, $\cpu$ in hours). The total memory allocated for the computation of $\mathbf{G}_{c,l,b}$ and the memory needed to store them are reported.}
    \label{tab:cyga}
\end{table}

%% file: fig_cyga_exp2_ch10.tex
\begin{figure*}
  \centering
  \includegraphics[height=0.28\textheight,keepaspectratio, trim=0cm 2cm 0cm 0.8cm, clip]{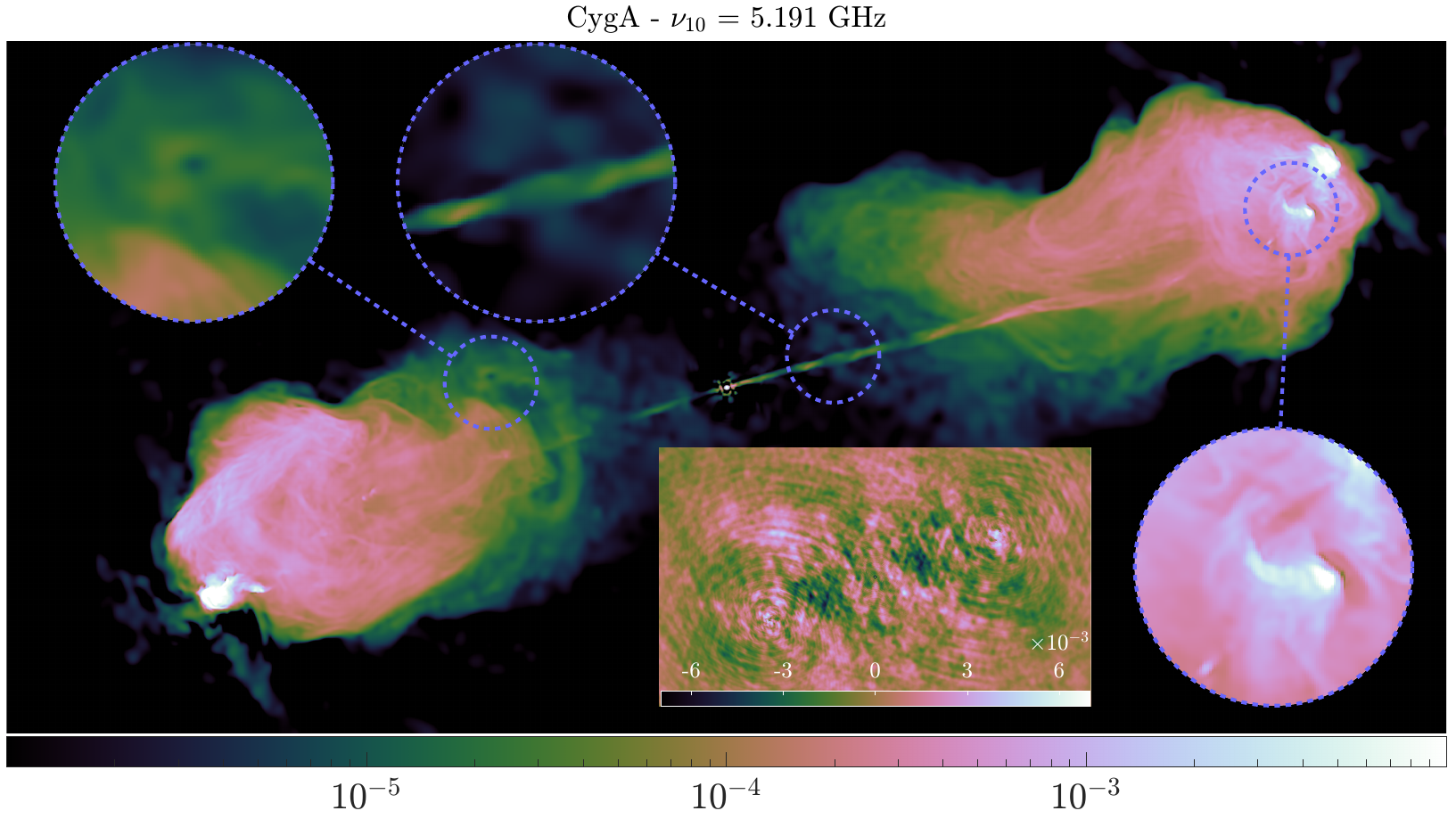}

  \vspace*{0.2cm}
  \includegraphics[height=0.28\textheight,keepaspectratio,trim=0cm 2cm 0cm 0.8cm, clip]{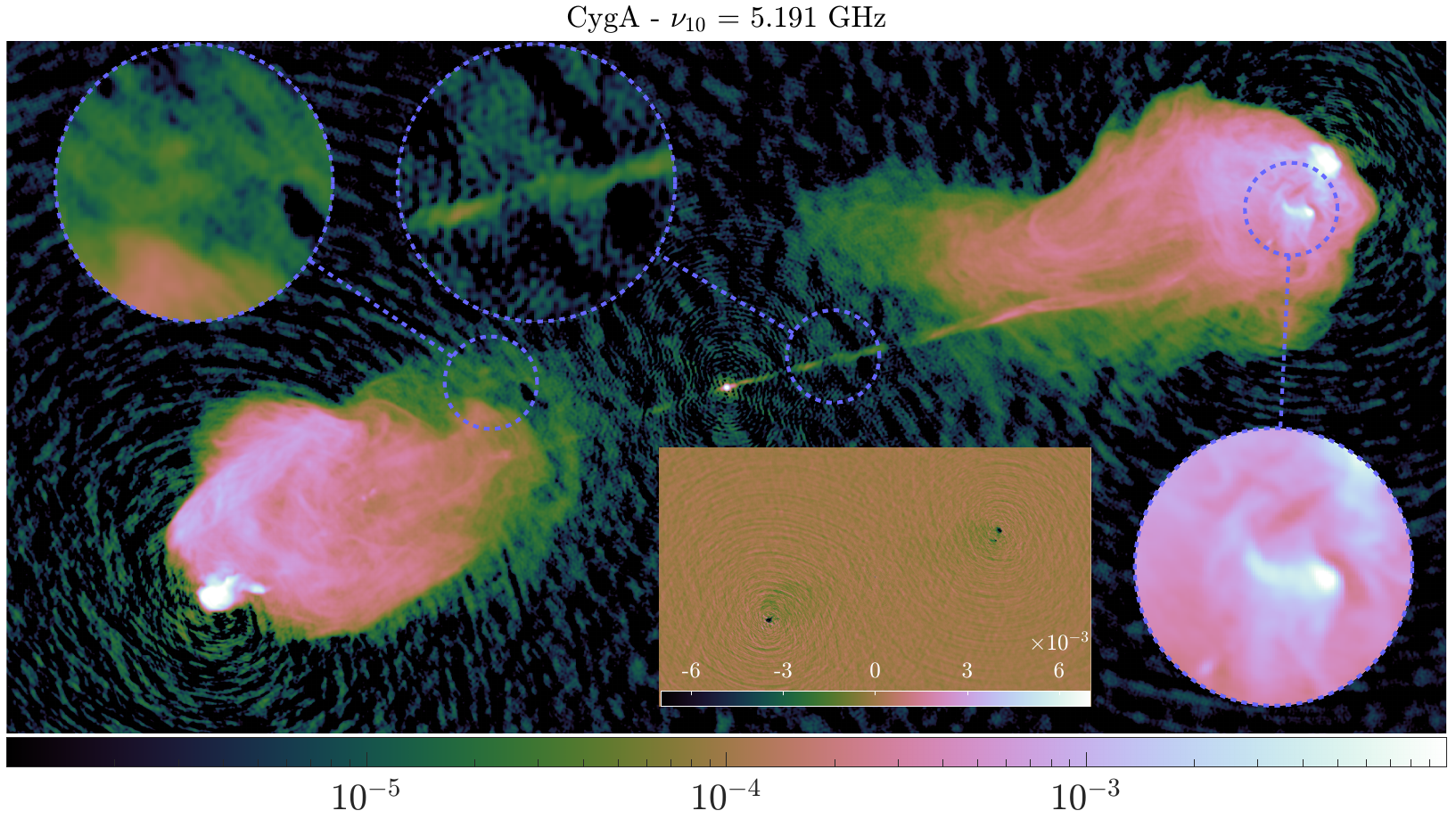}

  \vspace*{0.2cm}
  \includegraphics[height=0.325\textheight,keepaspectratio,trim=0cm 0cm 0cm 0.8cm, clip]{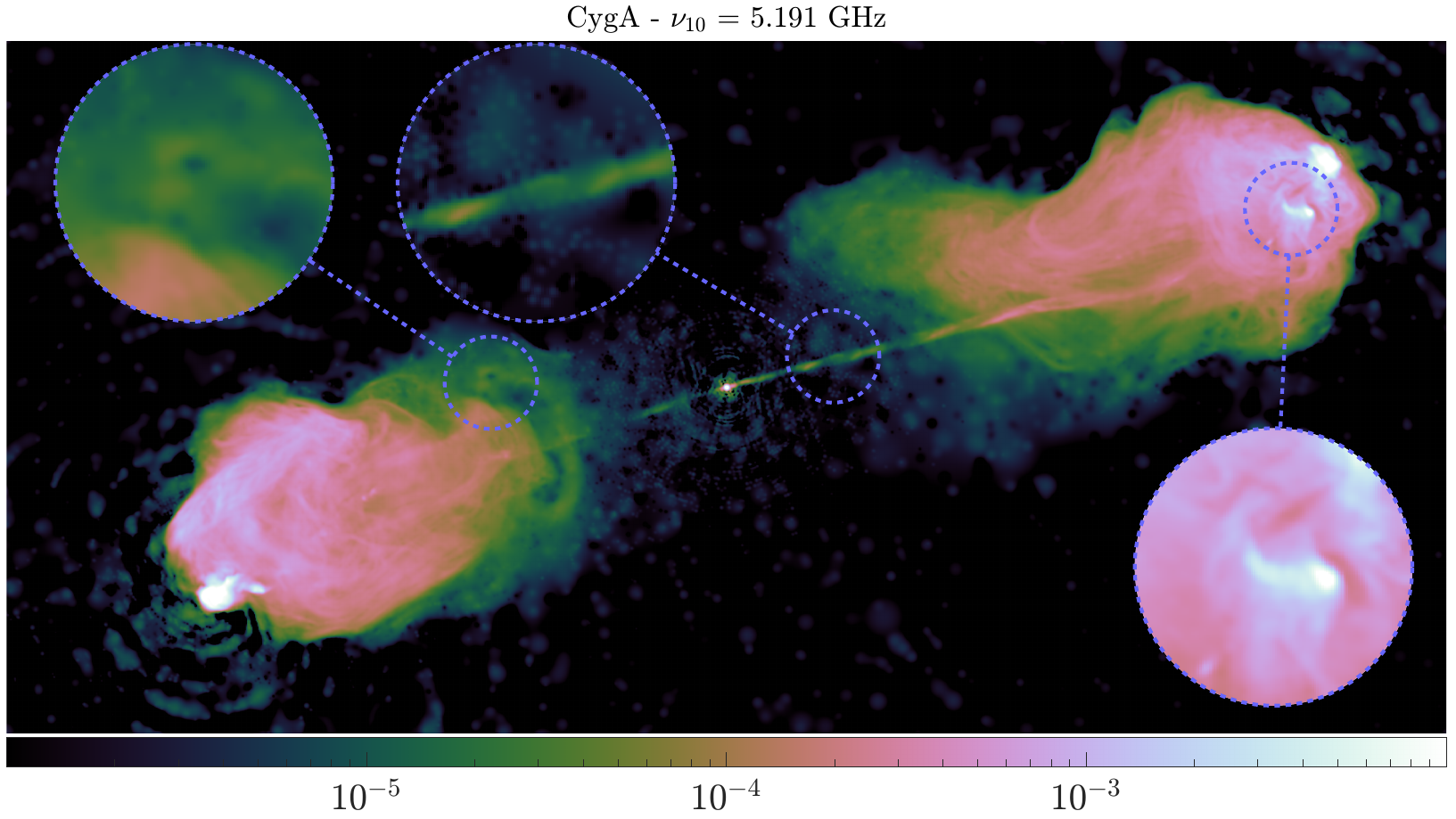}

  \caption{Experiment 2: Imaging results of Cyg A at the spectral resolution $\delta \nu=128$ MHz and spatial resolution $\delta x=0.0424\arcsec$. Reconstructions of channel $\widehat{\nu}_{10} = 5.191$ GHz. From top to bottom: estimated model image of Faceted HyperSARA, the restored image of the wideband CLEAN in WSClean, and the associated model image convolved with a restoring beam, obtained after the subtraction of the residual image, all in units of Jy/pixel and displayed in $\log_{10}$ scale (for visualisation purposes, WSClean images are normalized with the flux of the associated restoring beam). The images are overlaid with zooms on several regions of the radio galaxy, highlighted in dashed circles. Residual images of Faceted HyperSARA and WSClean, displayed in linear scale, are overlaying the respective model image (top row) and restored image (middle row). Their respective standard deviations, after a normalization by the flux of the associated restoring beams, are $0.006 $~mJy/pixel and $0.011$~mJy/pixel. Full image cubes are available online~\citep{faceting}.
  }
  \label{fig:cyga_dr_ch10}
\end{figure*}

%% file: fig_cyga_exp2_ch29.tex
\begin{figure*}
  \centering

  \includegraphics[height=0.28\textheight,keepaspectratio, trim=0cm 2cm 0cm 0.8cm, clip]{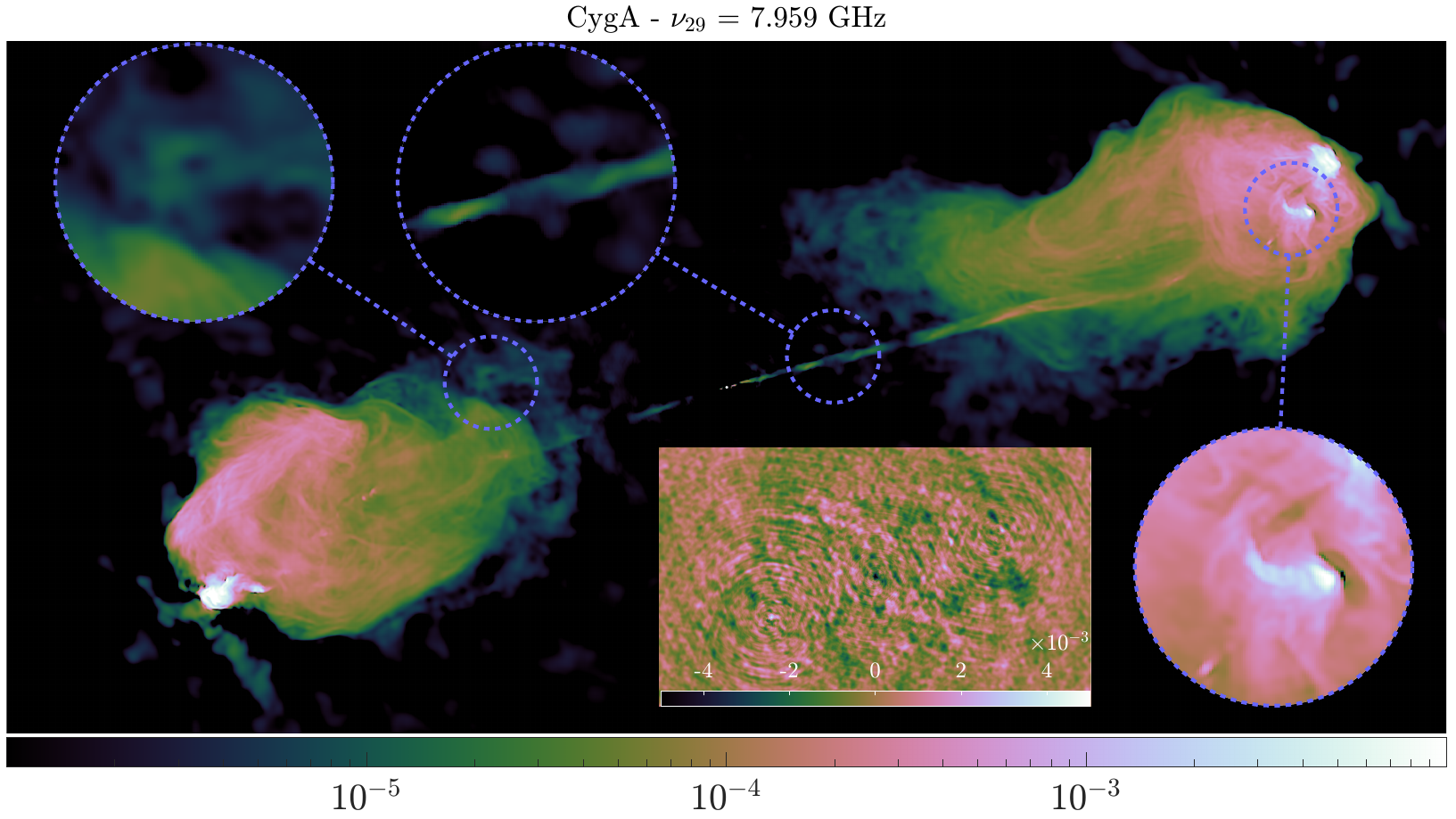}

  \vspace*{0.2cm}
  \includegraphics[height=0.28\textheight,keepaspectratio,trim=0cm 2cm 0cm 0.8cm, clip]{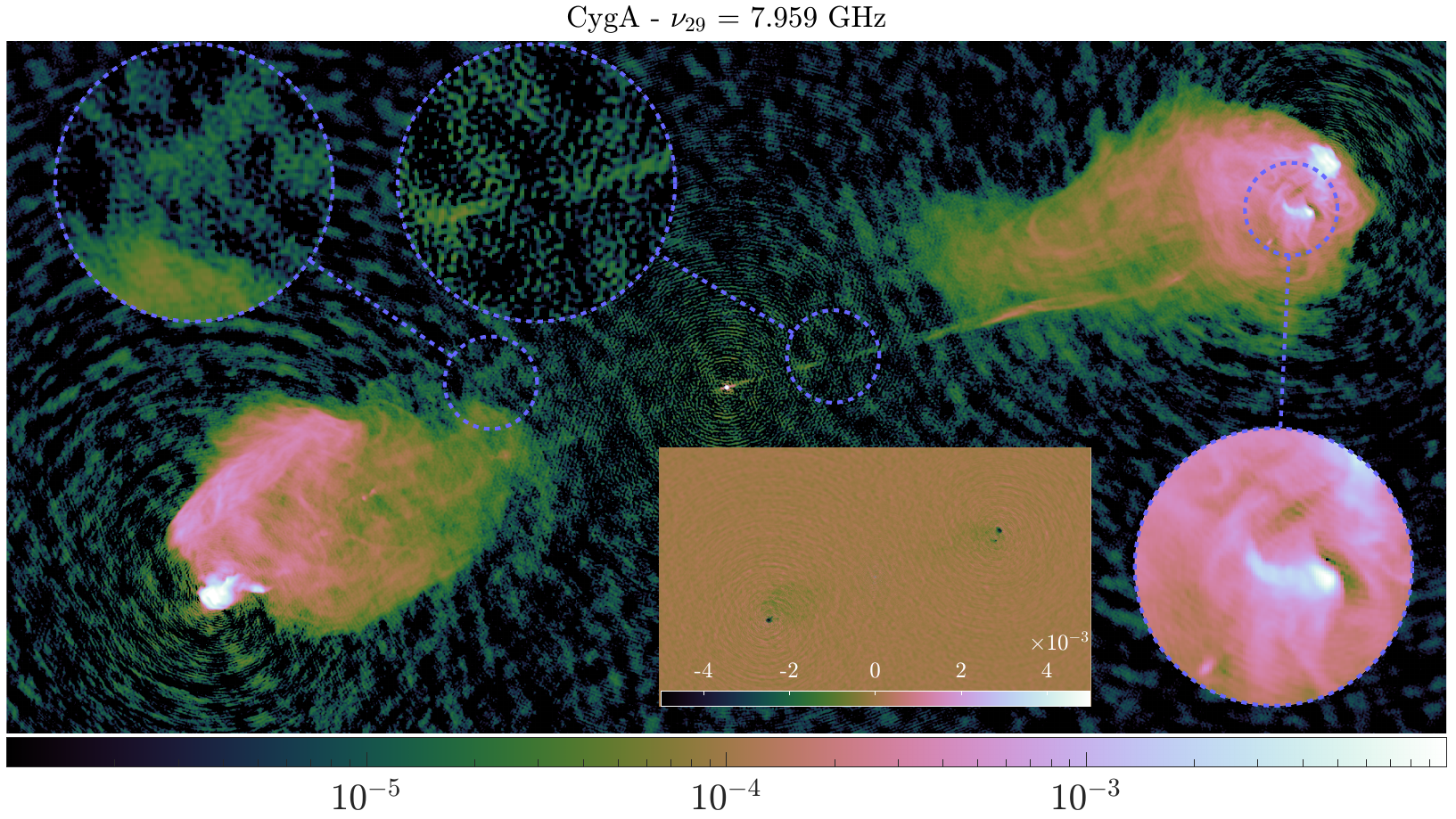}

  \vspace*{0.2cm}
  \includegraphics[height=0.325\textheight,keepaspectratio,trim=0cm 0cm 0cm 0.8cm, clip]{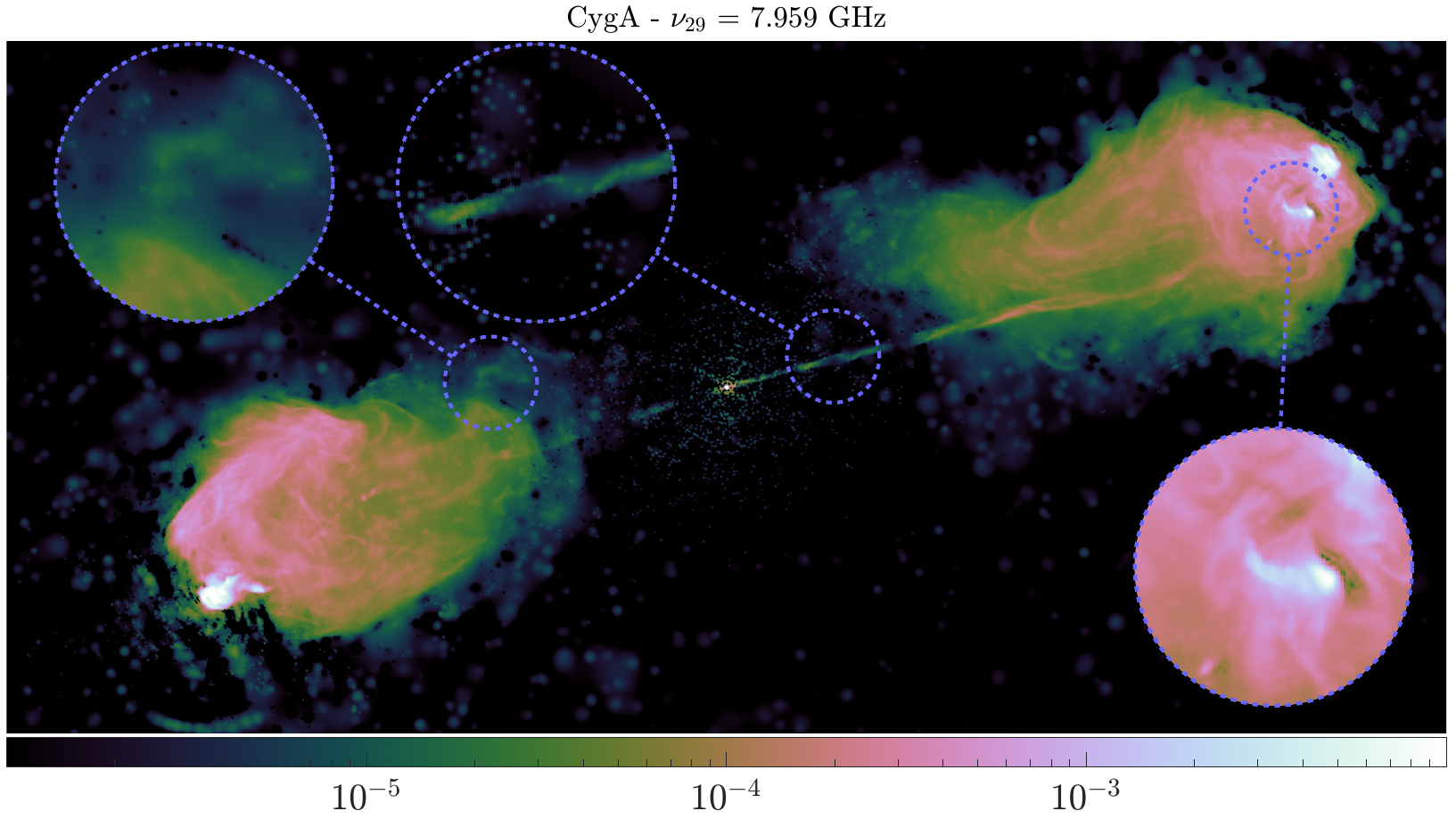}

  \caption{Experiment 2: Imaging results of Cyg A at the spectral resolution $\delta \nu=128$ MHz and spatial resolution $\delta x=0.0424\arcsec$. Reconstructions of channel $\widehat{\nu}_{29} = 7.959$ GHz. From top to bottom: estimated model image of Faceted HyperSARA, the restored image of the wideband CLEAN in WSClean, and the associated model image convolved with a restoring beam, obtained after the subtraction of the residual image, all in units of Jy/pixel and displayed in $\log_{10}$ scale (for visualisation purposes, WSClean images are normalized with the flux of the associated restoring beam). The images are overlaid with zooms on several regions of the radio galaxy, highlighted in dashed circles. Residual images of Faceted HyperSARA and WSClean displayed in linear scale are overlaying the respective model image (top row) and restored image (middle row). Their respective standard deviations, after a normalization by the flux of the associated restoring beams, are about $0.008$~mJy/pixel and $0.015$~mJy/pixel. Full image cubes are available online~\citep{faceting}.}
  \label{fig:cyga_dr_ch29}
\end{figure*}

%% file: table_timing_cyga_dr.tex
\begin{table}
    \centering
    {
    \begin{tabular}{ll} \toprule
        &Faceted HyperSARA \\ \midrule
        CPU cores   & 90 \\
        PDFB iterations & 2090 \\
        $\runp$ (s) & 55.57 $(\pm 3.70)$  \\ 
        $\run$ (h)  & 32.26  \\
        $\cpup$ (s) & 5001.3 $(\pm 333)$ \\
        $\cpu$ (h)  & 2903 \\
        Memory allocated to compute &  \multirow{2}{*}{203~GB} \\
        $\bs{\Lambda}_{l,b}\mathbf{H}_{l,b}$ matrices &  \\
        Memory occupied by &  \multirow{2}{*}{108~GB}  \\
        $\bs{\Lambda}_{l,b}\mathbf{H}_{l,b}$ matrices \\
        \bottomrule
    \end{tabular}}
    \caption{Experiment 2: Computing cost of Cyg A imaging at the spectral resolution $\delta \nu=128$ MHz and spatial resolution $\delta x=0.0424\arcsec$. Results are reported for Faceted HyperSARA ($C=1$, $Q=60$) in terms of number of CPU cores, number of PDFB iterations, runtime per iteration and total runtime ($\runp$ in seconds with the associated standard deviation, $\run$ in hours), CPU time per iteration and total CPU time ($\cpup$ in seconds with the associated standard deviation,
    $\cpu$ in hours). The total memory allocated for the computation of the holographic matrices and the the memory needed to store them are reported.
    }
    \label{tab:cyga_dr}
\end{table}